\documentclass[12pt,indent]{article}
\usepackage{a4,latexsym,amsmath}
\usepackage[latin1]{inputenc}
\usepackage{float}
\usepackage{natbib}
\usepackage{graphics}
\usepackage[english]{babel}
\usepackage{tabularx}
\usepackage{lscape}
\usepackage{multirow}
\usepackage{rotating}
\usepackage{dsfont}
\usepackage{subfig}
\usepackage{bbm}
\usepackage{booktabs}
\usepackage[table]{xcolor}

\usepackage{calc}
\usepackage{ifthen}
\newenvironment{tabularsmall}
{ \footnotesize \sffamily \tabular } {
\endtabular
\normalfont }

\newcommand{\deltab}{\boldsymbol{\delta}}

\newcommand{\xb}{\boldsymbol{x}}

\newcommand{\blanco}[1]{}

\def\d{\displaystyle}

\usepackage{natbib}

\usepackage{setspace}

\begin{document}
\bibliographystyle{chicago}
\sloppy

\makeatletter
\renewcommand{\section}{\@startsection{section}{1}{\z@}%
        {-3.5ex \@plus -1ex \@minus -.2ex}%
        {1.5ex \@plus.2ex}%
        {\reset@font\Large\sffamily}}
\renewcommand{\subsection}{\@startsection{subsection}{1}{\z@}%
        {-3.25ex \@plus -1ex \@minus -.2ex}%
        {1.1ex \@plus.2ex}%
        {\reset@font\large\sffamily\flushleft}}
\renewcommand{\subsubsection}{\@startsection{subsubsection}{1}{\z@}%
        {-3.25ex \@plus -1ex \@minus -.2ex}%
        {1.1ex \@plus.2ex}%
        {\reset@font\normalsize\sffamily\flushleft}}
\makeatother



\newsavebox{\tempbox}
\newlength{\linelength}
\setlength{\linelength}{\linewidth-10mm} \makeatletter
\renewcommand{\@makecaption}[2]
{
  \renewcommand{\baselinestretch}{1.1} \normalsize\small
  \vspace{5mm}
  \sbox{\tempbox}{#1: #2}
  \ifthenelse{\lengthtest{\wd\tempbox>\linelength}}
  {\noindent\hspace*{4mm}\parbox{\linewidth-10mm}{\sc#1: \sl#2\par}}
  {\begin{center}\sc#1: \sl#2\par\end{center}}
}



\def\R{\mathchoice{ \hbox{${\rm I}\!{\rm R}$} }
                   { \hbox{${\rm I}\!{\rm R}$} }
                   { \hbox{$ \scriptstyle  {\rm I}\!{\rm R}$} }
                   { \hbox{$ \scriptscriptstyle  {\rm I}\!{\rm R}$} }  }

\def\N{\mathchoice{ \hbox{${\rm I}\!{\rm N}$} }
                   { \hbox{${\rm I}\!{\rm N}$} }
                   { \hbox{$ \scriptstyle  {\rm I}\!{\rm N}$} }
                   { \hbox{$ \scriptscriptstyle  {\rm I}\!{\rm N}$} }  }

\def\d{\displaystyle}

\title{Item focussed  Trees for the Identification of Items in  Differential Item Functioning }
\author{Gerhard Tutz \& Moritz Berger\\{\small Ludwig-Maximilians-Universit\"{a}t M\"{u}nchen}\\
{\small Akademiestra{\ss}e 1, 80799 M\"{u}nchen}}


\maketitle

\begin{abstract} 
\noindent
A new method for the identification of differential item functioning (DIF) by using recursive partitioning techniques is proposed. We assume an extension of the Rasch model that allows for DIF being induced by an arbitrary number of covariates for each item. Recursive partitioning on the item level results in one tree for each item and leads to simultaneous selection of items and variables that induce DIF.  For each item it is possible to detect groups of subjects with different item difficulties, defined by combinations of characteristics that are not pre-specified. An algorithm is proposed that is based on permutation tests. Various simulation studies, including the comparison with traditional approaches to identify items with DIF, show the applicability and the competitive performance of the method. Two applications illustrate the usefulness and the advantages of the new method.
\end{abstract}


\section{Introduction}\label{sec:introduction}

Differential item functioning (DIF) is a well known problem in item response theory. It occurs if
the probability of a correct response among equally able
persons differs in subgroups, for example, if the difficulty of an item
 depends on the membership to a  racial, ethnic or gender
subgroup. Then the performance of a  group can be lower because
these items are related to specific knowledge that is less present
in this group.  The effect is measurement bias and possibly
discrimination, see, for example,
\citet{millsap1993methodology}, \citet{zumbo1999handbook}. Various forms of
differential item functioning have been considered in the
literature, see, for example, \citet{wainer1993differential,osterlind2009differential,rogers2005differential}.  In particular \citet{magisetal:2010} gave an excellent overview of the existing DIF detection methods.

The traditional approach to identify items that carry DIF is based on test statistics. For each item a test is performed that shows if the item has different difficulties in subgroups that have to be defined by the experimenter. Test statistics  have been proposed by  \citet{thissen1993detection}, \citet{Lord:1980}, \citet{holland1988}, \citet{kim1995} and \citet{raju1988area}. Mixed model approaches were proposed by \citet{van2005assessing} and Bayesian approaches have been developed by \citet{soares2009integrated}.

The testing approach is not without problems. First, when testing it is assumed that all other items are free of DIF, called item purification, which is an assumption that typically does not hold, see also \citet{magisetal:2010}.
Second, the proposed tests are limited to the consideration of few subgroups. Typically one considers just two subgroups  with one group being fixed as the reference group. That means if one suspects item difficulties  to depend on age one has to know the age groups before testing. Thus age has to be split into two or more intervals without knowing which ones are relevant. Moreover, the approaches are restricted to subgroups. Therefore, it is hard to investigate the dependence on more than one possibly DIF inducing variable.

More recently, several methods have been proposed to cope with these problems. \citet{TuSchauDIFPsych} proposed  an explicit model for differential item functioning that  includes a set of variables, containing metric as well as categorical components, as potential candidates for inducing
DIF. The abundance of parameters in the model is handled by using  penalization techniques.
The procedure allows to identify the items that suffer from DIF and investigate which
variables are responsible. An alternative  approach  that is also able to handle several groups and continuous
variables was proposed by \citet{Stretal:2013:raschtree}. It avoids the comparison of  pre-specified focal and reference group by using recursive partitioning techniques, also known as trees. The proposed recursive partitioning scheme automatically identifies  the subgroups of subjects exhibiting DIF.

The method proposed here also uses recursive partitioning techniques, but in a different form than \citet{Stretal:2013:raschtree}.
\citet{Stretal:2013:raschtree} recursively partition the covariate space
to identify regions of the covariate space in which DIF occurs. In the investigated regions a parametric latent trait model that includes covariates is fitted. Regions are suspected to be relevant if the parameter estimates in the regions differ strongly. Therefore, regions in the covariate space are identified that show different difficulties. A disadvantage of the method is that it detects regions of the covariate space that are linked to DIF but does not automatically detect the items that are responsible.
In contrast, the recursive partitioning method proposed here focusses on the detection of the items that are responsible for DIF. Recursive partitioning is used on the item level not on the global level, which treats all items simultaneously, as in the method proposed by \citet{Stretal:2013:raschtree}. The item-focussed approach allows to detect the items that carry DIF but keeps the advantage that no pre-specified subgroups are needed.

In Section \ref{sec:method} we introduce the new method and present an illustrative example, in Section
\ref{sec:basic_algorithm} we give a detailed description of the fitting procedure. Results of wider simulations studies are given in Section \ref{sec:simulation}, in Section \ref{sec:application} we consider another application.

\section{Item Focussed Recursive Partitioning} \label{sec:method}

We will consider differential item functioning for the Rasch model. Therefore we start with the introduction of some notation.

\subsection{Differential Item Functioning for the Rasch Model}

In the binary  Rasch model  the probability for a person to score on an item is determined by a parameter for the latent ability of the person and a parameter for the item difficulty. In the case of $P$ persons and $I$ items, the Rasch model is given by

\begin{equation}
\label{RM1}
P(Y_{pi}=1)
 =\dfrac {\exp(\theta_p -\beta_i)} {1+\exp(\theta_p -\beta_i)} \quad
 p=1,\dots,P~~~,i=1,\dots,I,
\end{equation}
 where $Y_{pi}$ represents the response of person $p$ on item $i$. It is coded by $Y_{pi}=1$ if person $p$ solves item $i$ and $Y_{pi}=0$ otherwise. Both the person parameters, $\theta_p,~p=1,\ldots,P$, and the item parameters, $\beta_i,~i=1,\ldots, I$, are unknown and have  to be estimated.

An alternative form of the model is

\begin{equation}
\label{RM2}
\log \left(\dfrac{P(Y_{pi}=1)}{P(Y_{pi}=0)}\right)= \eta_{pi}=\theta_p - \beta_i,
\end{equation}
where  the predictor $\eta_{pi}=\theta_p - \beta_i$ represents the difference between ability of the person and difficulty of the item. As model \eqref{RM2} is not identifiable in this general form, a restriction on the parameters is needed. A common choice that is also used in the following is $\theta_P=0$.

In item response models, DIF appears if an item has different difficulties depending on characteristics of the person which tries to solve the item.  The simplest form of DIF is found if items difficulties differ in a focal and a reference group. If item $i$ is a DIF item the predictor is given by

\begin{equation}
\label{DIFl}
\eta_{pi}=\theta_p -\gamma_i^{(j)}, j=1,2,
\end{equation}
where $j=1$ denotes the focal group and $j=2$ the reference group. DIF occurs, if $\gamma_i^{(1)} \ne \gamma_i^{(2)}$, which can be tested, for example, by likelihood ratio tests.
The recursive partitioning scheme considered in the following  uses this simple model, which considers two subgroups, as building block. By iterative application of the splitting into two subgroups
one obtains a tree for each item.

\subsection{Recursive Partitioning}

Recursive partitioning also known as tree-based modeling
has its roots in automatic interaction detection (AID), proposed by
\citet{MorSon:63}. The most popular modern version is due to
\citet{BreiFrieOls:84} and is known by the name
\textit{classification and regression trees}, often abbreviated as
CART. Alternative approaches are the C4.5 algorithm (\citet{Quinlan:86},
\citet{Quinlan:93}) or the recursive partioning framework based on conditional inference proposed by \citet{Hotetal:2006}. The method is
conceptually very simple. By binary recursive partitioning
 the feature space is partitioned
into a set of rectangles, and on each rectangle a simple model (for
example, a constant) is fitted. An overview with a focus on psychometrics was given by \citet{Strobetal:2009}.

In regression  trees may be seen as a hierarchical way to describe a
partition of the predictor space. The  tree represents  the partition in a unique way.
Each node of the tree corresponds to a subset of the predictor space.
The \textit{root}  is the top node  consisting of
the whole predictor space, and the \textit{terminal nodes}  or \textit{leaves} of the tree correspond to the
subregions.

To grow a tree one typically uses the
\grqq standard splits\grqq, which means that each
partition of node $A$ into subsets $A_1, A_2$ is determined by only
one variable. The splits to be considered depend on the scale of the
variable:
\begin{itemize}
\item[] For \textit{metrically scaled} and \textit{ordinal} variables, the partition into two subsets has the form
\[
A \cap \{ x_i \leq c \}, \quad A \cap \{ x_i > c \},
\]
based on the threshold $c$ on variable $x_i$.
\item[] For \textit{categorical} variables without ordering $x_i \in \{1,\ldots,k_i\}$, the partition has the form
\[
A \cap S, \quad A \cap \bar S,
\]
where $S$ is a non-empty subset $S\subset\{1,\ldots,k_i\}$ and $\bar
S=\{1,\ldots,k_i\}\setminus S$ is the complement.
\end{itemize}

In the following we will mostly use the split for metrically scaled or ordinal variables to illustrate how trees are obtained.
Let  $\boldsymbol{x}_p^T=(x_{p1},\ldots,x_{pm})$ denotes a person-specific covariate vector of length $m$.
For the detection of DIF the first split means one examines for all the items, all the variables and possible  splits of the corresponding variable
the Rasch model with predictor
\[
\eta_{pi} = \theta_p - [\gamma_{il} I(x_{pj} \le c_{j})+ \gamma_{ir} I(x_{pj} > c_{j})],
\]
where $I(.)$ denotes the indicator function with $I(a)=1$ if $a$ is true and $I(a)=0$ otherwise.
The model is just an alternative representation of (\ref{DIFl}), with the focal and  reference group constructed by a split of the $j$th variable at split-point  $c_{j}$.
The parameter $\gamma_{il}$ denotes the item difficulty in the left node ($x_{pj} \le c_{j}$) and  $\gamma_{ir}$ the item difficulty in the right node ($x_{pj} > c_{j}$).
One chooses that combination of item, variable and split that has the smallest $p$-value when tested for DIF, that is, in the examination of the null hypothesis
$H_0: \gamma_{il}-\gamma_{ir}=0$. This selection yields the first split into left and right daughter nodes corresponding to the regions $I(x_{pj} \le c_{j})$ and $I(x_{pj} > c_{j})$.

Further splitting means that one of the nodes, say $I(x_{pj} > c_{j})$,  is further split, for example, in variable $s$ at cut point $c_{s}$, yielding the daughters
\[
I(x_{pj} > c_{j})I(x_{ps} \le c_{s}) \quad \text{and}  \quad I(x_{pj} > c_{j})I(x_{ps} > c_{s}).
\]
and the linear predictor
\[
\eta_{pi} = \theta_p - [\gamma_{il} I(x_{pj} \le c_{j})+ \gamma_{il}^{(n)} I(x_{pj} > c_{j})I(x_{ps} \le c_{s})+\gamma_{ir}^{(n)}I(x_{pj} > c_{j})I(x_{ps} > c_{s})] ,
\]
where $\gamma_{il}^{(n)}, \gamma_{ir}^{(n)}$ are the weights on the new split.
Then the item difficulty in the region $\{x_{pj} \le c_{j}\}$ is $\gamma_{il}$ but for the region $\{x_{pj} > c_{j}\}$ one has to distinguish between $\{x_{pj} > c_{j},x_{ps} \le c_{s}\}$ with item difficulty $\gamma_{il}^{(n)}$ and $\{x_{pj} > c_{j},x_{ps} > c_{s}\}$ with item difficulty $\gamma_{ir}^{(n)}$.


The corresponding trees are trees for specific items, namely the items that were selected to carry DIF. If an item is never selected it is considered as compatible with the Rasch model.

\begin{figure}[!ht]
\centering
\includegraphics[width=1\textwidth]{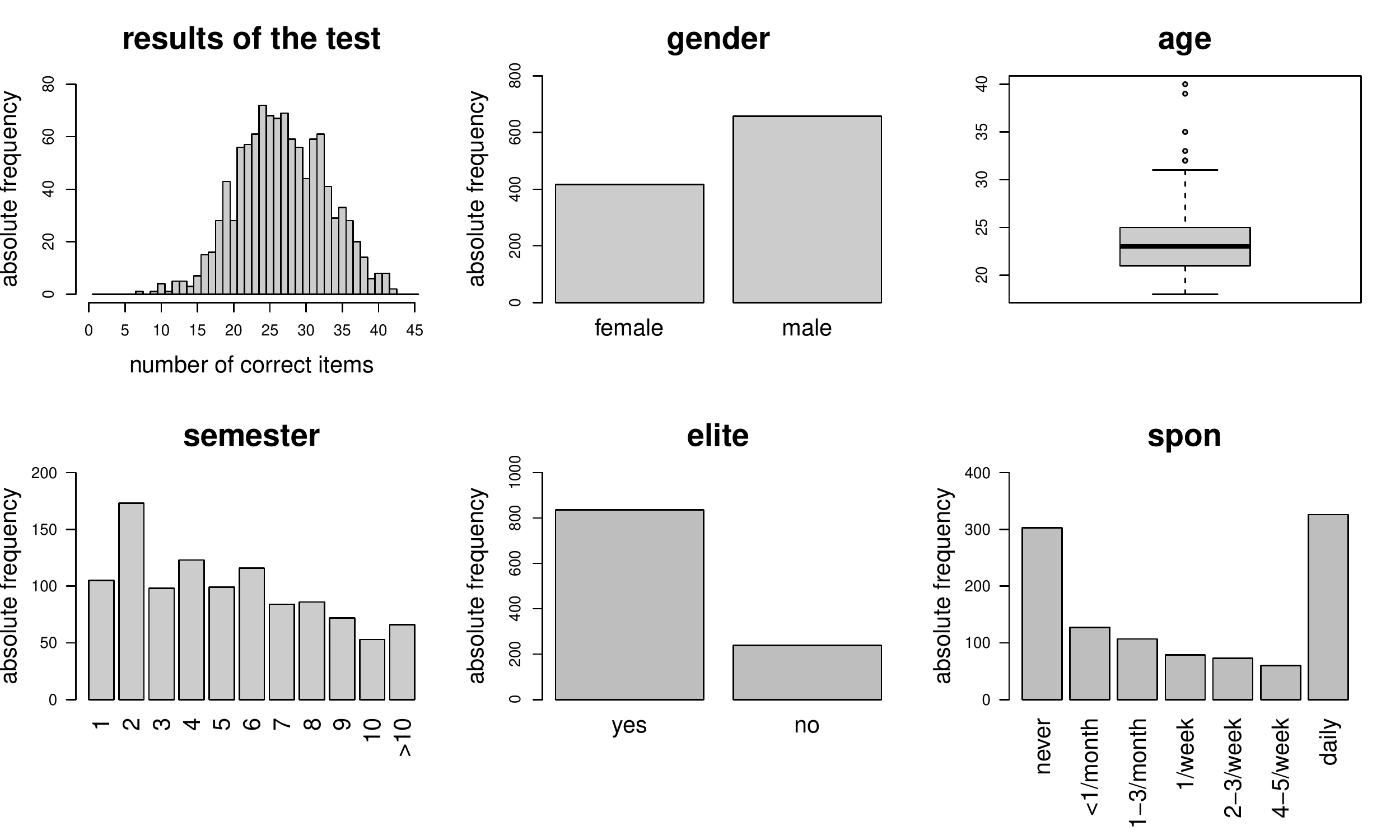}
\caption{Graphical representation of the results of the online quiz (upper left) and the distribution of the five covariates in the analyzed data.}
\label{fig:Deskriptiv_SPISA}
\end{figure}

\subsection{An Illustrative Example}
Before giving the details how to grow trees we want to illustrate the procedure by use of a data set that has been used previously in the DIF literature \citep{Stretal:2013:raschtree}.  We consider the data of an online quiz testing one's general knowledge. The test was conducted by the German news magazin Spiegel in 2009. The whole test consisted of 45 questions from five different topics, that are politics, history, economy, culture and natural sciences. A detailed analysis and discussion of the original data set is found in \citet{Trepte2010SPISA}.

We use a subset of the data including 1075 university students from Bavaria. To test for DIF we incorporate the five covariates gender (0:female, 1:male), age, number of matriculated semester, elite status of the university (0:no, 1:yes) and the frequency of accessing Spiegel's online magazine (spon) from 1 (never) to 6 (daily).
The distributions of the five covariates and the test results are displayed in Figure \ref{fig:Deskriptiv_SPISA}.

\begin{figure}[!ht]
\begin{center}
\includegraphics[width=0.8\textwidth]{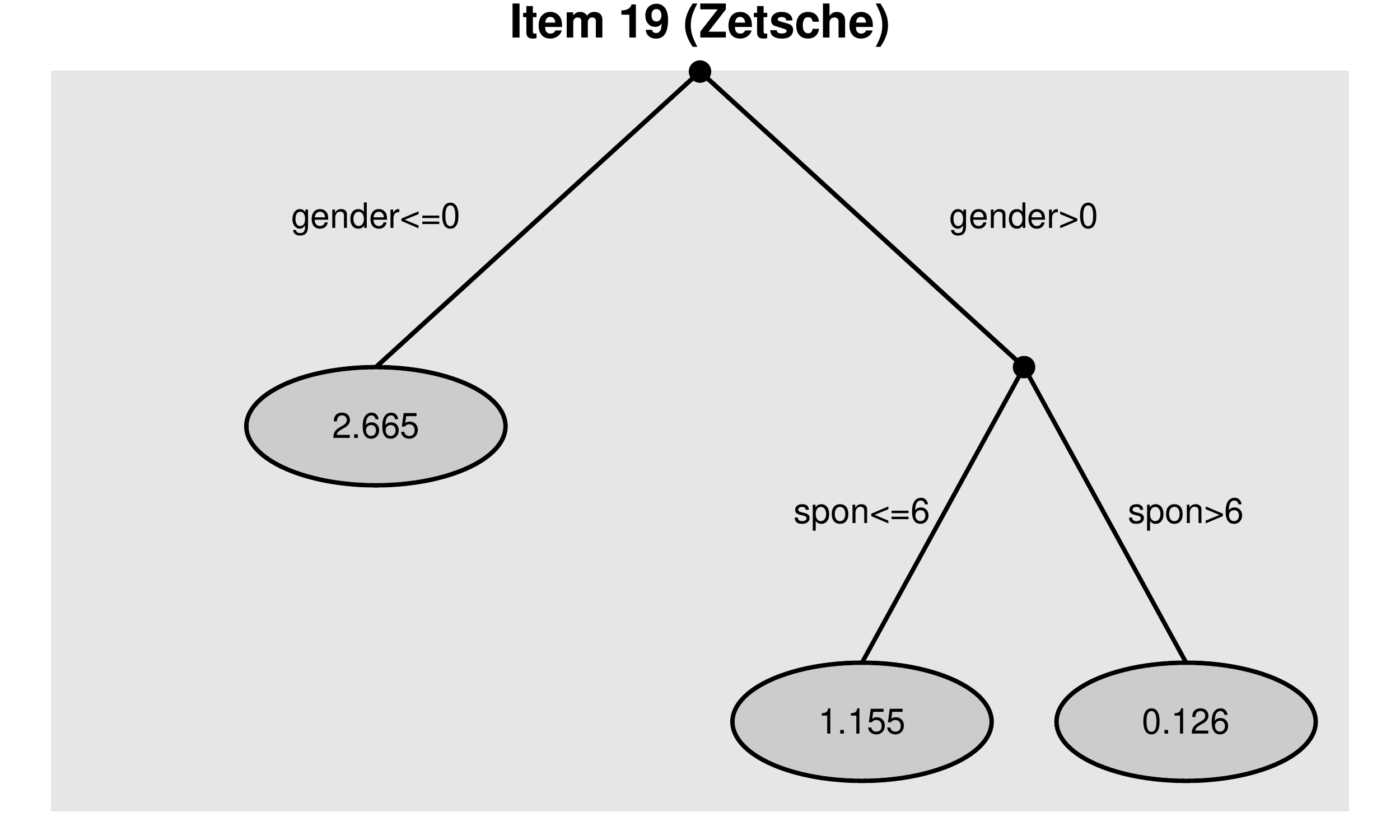}
\includegraphics[width=0.4\textwidth]{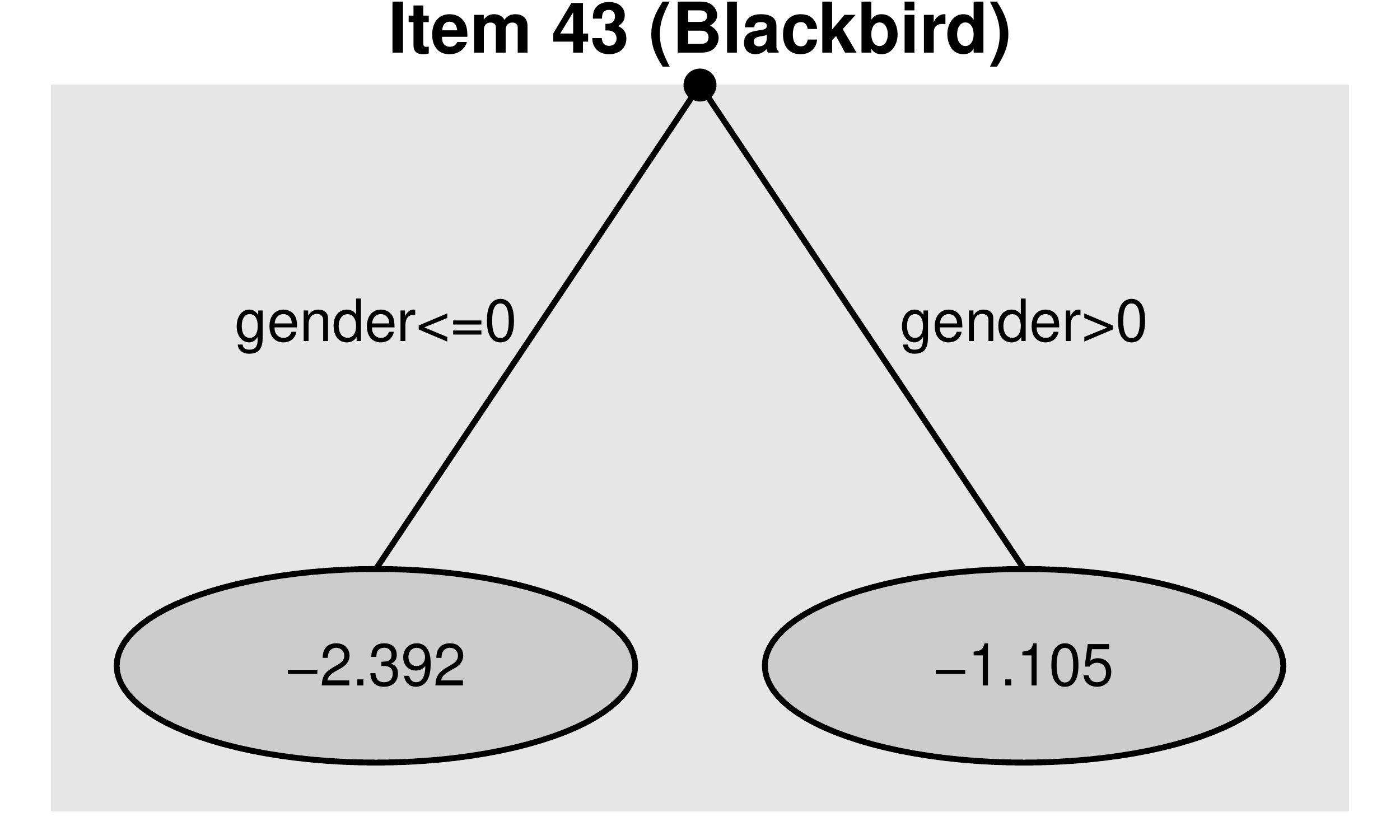}
\includegraphics[width=0.4\textwidth]{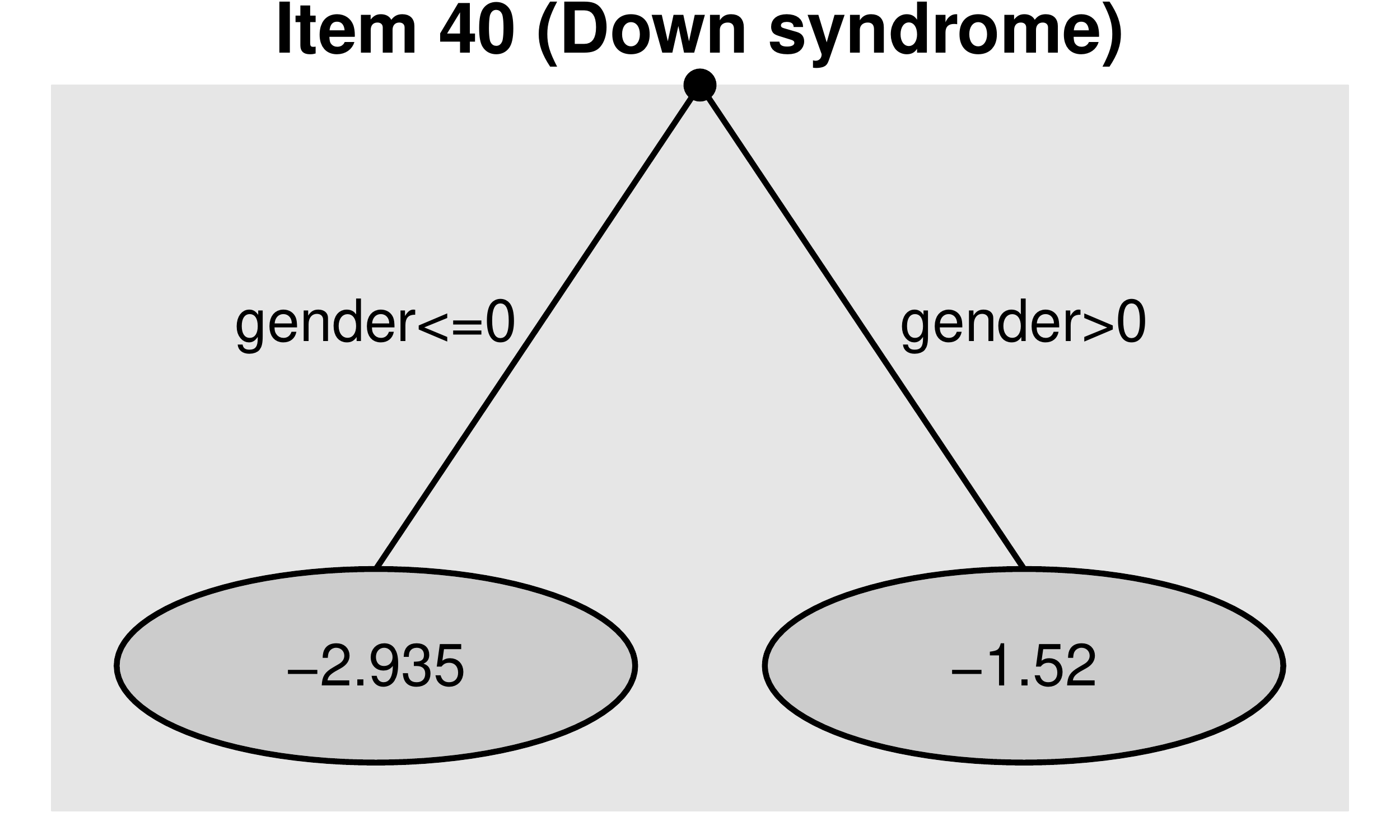}
\end{center}
\caption{Trees for Items 19, 43 and 40 of the general knowledge test. The parameters of item-difficulty are given for each subgroup represented by the leaves of the trees.}
\label{fig:Tree_194340}
\end{figure}

\subsubsection*{Item Focussed Recursive Partitioning}

When using item focussed recursive partitioning  21 of the 45 items  show DIF. The result is not surprising because the questions of the quiz were not chosen very carefully to avoid DIF. Altogether the algorithm performs 33 splits until  further splits are not significant at significance level $\alpha=0.05$  (for details of the test see Section \ref{sec:basic_algorithm}). The first ten splits all refer to the covariate gender, so the strongest effects were found for the difference between males and females. No significant splits were found for the variable elite. The difficulties of the items seem not to depend on the  elite status of the university. The three items with the strongest effects, which were found in the first iterations of the algorithm, were the following:

\begin{itemize}
\item[] 19: Who is this? - Picture of Dieter Zetsche, CEO of Mercedes-Benz
\item[] 43: Which kind of bird is this? - Blackbird
\item[] 40: What is also termed Trisomy 21? - Down syndrome
\end{itemize}

The resulting trees for these items 19, 43 and 40 are shown in Figure \ref{fig:Tree_194340}. For each item one can see how the difficulty of the item depends on the characteristic of certain variables. The estimated  item difficulties are given in each leaf of the trees, which  represent the identified subgroups. For  example, in item 19 (recognition  of Dieter Zetsche as CEO of Mercedes Benz) the difficulty for females (gender=0) is 2.665 while for males (gender=1) it is distinguished between students who rather frequently  read Spiegel online (spon $>$ 6) with an item difficulty of 0.126 and a much larger item difficulty of 1.155 for students who read it less regularly (spon $ \le 6$). The other two items show DIF only for gender. Both items concerning the recognition of birds and knowledge of genetic diseases are easier to solve for females. It is also seen that item 19 is much harder to solve than the other two items.

\begin{figure}[!ht]
\begin{center}
\includegraphics[width=0.9\textwidth]{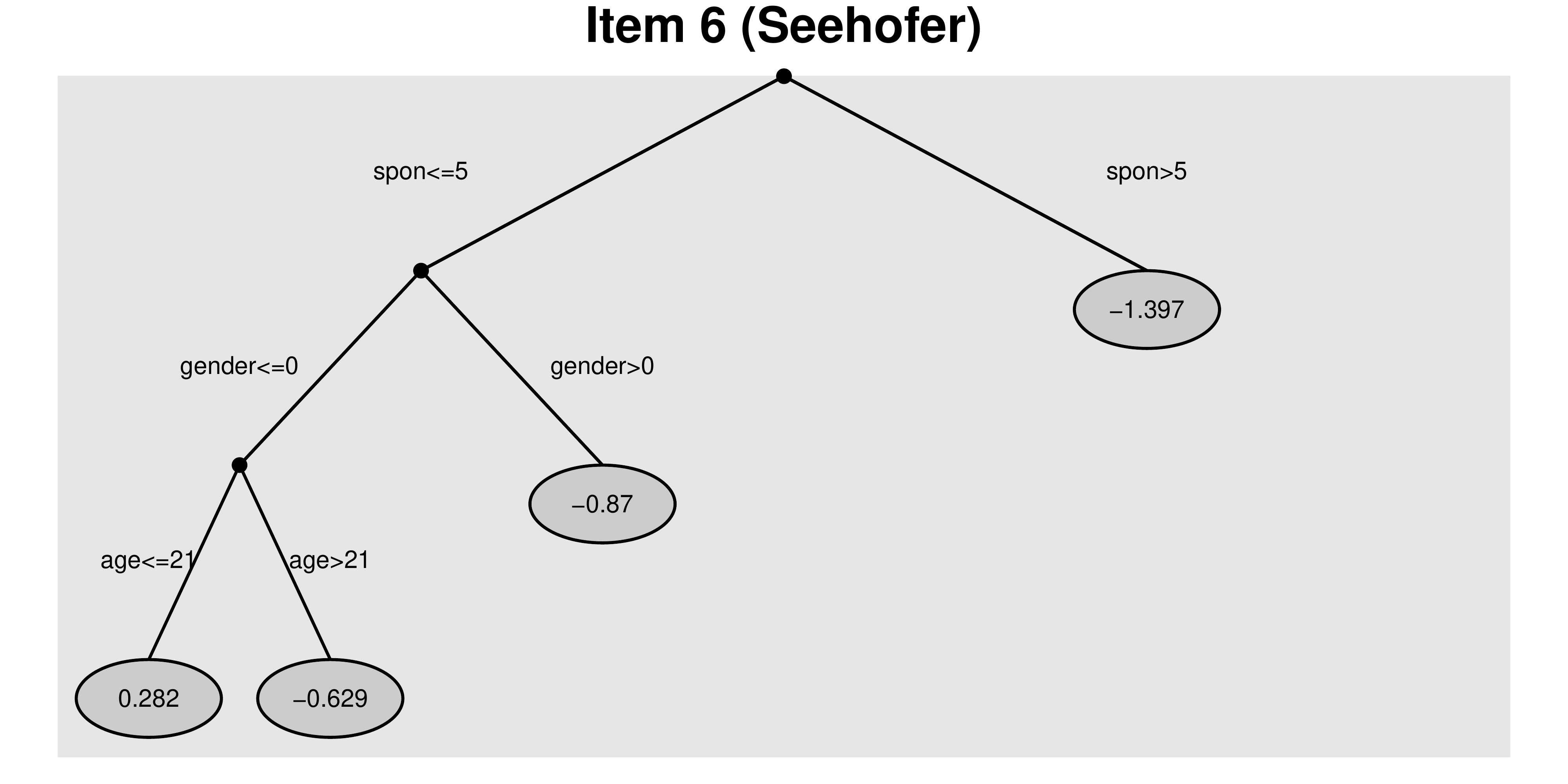}
\end{center}
\caption{Tree for Item 6 of the general knowledge test. Parameters of item-difficulty are given in each leaf of the tree.}
\label{fig:Tree_6}
\end{figure}

Another quite interesting tree structure is received for Item 6 of the test (see Figure \ref{fig:Tree_6}). The corresponding question asks to identify the Prime Minister of Bavaria, Horst Seehofer.
For all students who read the online magazine very regular (spon$>$5) the question  is very easy. By contrast the question is more difficult for students who do not read Spiegel online very often (spon $ \le 5$), in particular if they are female  (gender=0) and comparably young (age $\le 21$).

The strength of the approach is that one sees for each items which variables generate DIF.  The tree structure also yields an ordering of the relevance of the variables with the first split being the most relevant. By recursive splitting of regions trees are always devices to detect interactions. For example, in item 19 a relevant interaction effect is that of gender and frequency of reading Spiegel online.  Moreover, trees automatically detect the groups that have to be distinguished. It is not necessary to define the focus and the reference group beforehand.

\begin{figure}[!ht]
\begin{center}
\includegraphics[width=1\textwidth]{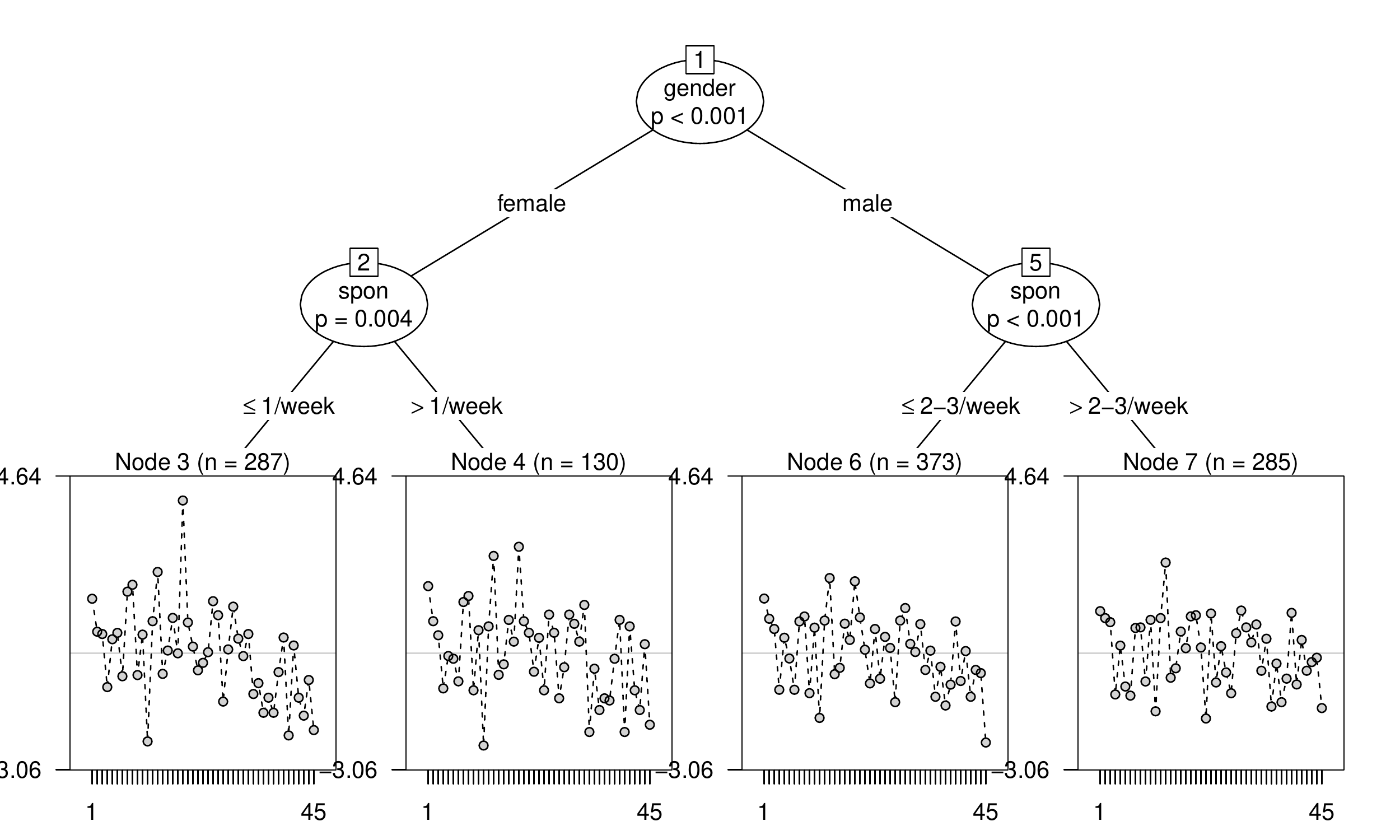}
\end{center}
\caption{Result of the analysis of the Spiegel online data by a Rasch tree.}
\label{fig:spisa_raschtree}
\end{figure}

\subsubsection*{Rasch Trees}
To illustrate the difference between the alternative approach to use trees we analyse in the following the same data set by using the Rasch tree concept of \citet{Stretal:2013:raschtree}. The corresponding tree is given in Figure \ref{fig:spisa_raschtree}. The significance level used for the tests for parameter instability was the same as for our tree,  $\alpha=0.05$.

The basic concept of conventional Rasch trees is to search for the split in the explanatory variables that shows the strongest differences in \textit{all of the item difficulties}. In this application one obtains a tree  with splits in two variables, gender and spon. These variables are found to induce DIF and one finds four groups that differ in terms of item difficulty. In each leaf of the corresponding tree the estimated difficulties are shown. The crucial point is that the resulting tree is one tree for all of the items. It does not identify the items that are responsible for the split and therefore for DIF. Consequently it is hard  to identify those items that are affected by DIF and those that are not from Figure \ref{fig:spisa_raschtree}. Moreover, there is no criterion provided to identify the responsible items. In contrast, the  item focussed tree shows which items are responsible. It is seen from Figure \ref{fig:Tree_194340}  that both variables, gender and spon, are also found for items 19, 34 and 40 but in a more differentiated way. Figure \ref{fig:Tree_6} shows that also item 6 is a DIF item that is also specific for  age.

\section{Fitting Trees} \label{sec:basic_algorithm}
In this section we give the details of the algorithm that yields item focussed trees. In particular we show how trees are grown and when to stop.

\subsubsection*{The Basic Algorithm}

In all tree-based methods one has to decide in particular how to split and how to determine the size of the trees. Split criteria that are in common use are splitting by impurity measures like the Gini-based impurity or the entropy and  test-based splits. The latter  use a test statistic to evaluate which split is the strongest to explain the impact of predictors.
Already \citet{BreiFrieOls:84} considered very general families of impurity measures including the entropy, which is strongly related to test-based split when the deviance is used as test statistic, see, for example,  \citet{CiaChaHogMcK:87} and \citet{ClaPre:92}. As far as tree size is concerned,
in  early recursive partitioning approaches the final tree is typically obtained by growing large trees and then prune them to an adequate size, for details see
\citet{BreiFrieOls:84} or \citet{Ripley:96}, Chapter 7. Alternative methods are based on maximally selected statistics.
The basic idea  is to consider the distribution of the
selection process. When a split-point is selected based on a test
statistic  $T_i$ for  possible
split-point $i$, one investigates the distribution of
$T_{max}=max_{i=1,\dots,m} T_i$. The $p$-value of the distribution
of $T_{max}$ provides a measure for the relevance of a predictor
that does not depend on the number of split-points since the number
has been taken into account, see
 \citet{HotLau:03}, \citet{Shih:04}, \citet{ShiTsa:2004}, \citet{StrBouAug:2007}.
A unified framework for recursive partitioning that embeds tree-structured
regression models into a well-defined theory of conditional
inference procedures was proposed by \citet{Hotetal:2006}. The splitting is stopped when the global null
hypothesis of independence between the response and any of the
predictors cannot be rejected at a pre-specified nominal
significance level $\alpha$. The method explicitly accounts for the
involved multiple test problem. By separating variable selection and
splitting procedure one arrives at an unbiased recursive
partitioning scheme that also avoids the selection bias toward
predictors with many possible splits or missing values. We will draw on the concept of conditional inference procedures
in our approach to select splits.

Let us consider again the  construction of the first split. One examines for all the items, all the variables and possible  splits of the corresponding variable
the Rasch model with predictor
\[
\eta_{pi} = \theta_p - [\gamma_{il} I(x_{pj} \le c_{j})+ \gamma_{ir} I(x_{pj} > c_{j})],
\]
The test for DIF at split point $c_{j}$ corresponds to the null hypothesis
$H_0: \gamma_{il}-\gamma_{ir}=0$. If $H_0$ holds for all split points the item shows no DIF since $\gamma_{il}=\gamma_{ir}$ holds for all split points.
Let $T_{jc_j}$ denote the corresponding test statistic, for example, the log-likelihood test statistic.
To obtain a test for variable $j$ one has to consider simultaneously all the test statistics $T_{jc_j}$ with $c_j$ from the set of possible splits.
We will use the maximal value statistic $T_j= \max_{c_j} T_{jc_j}$, which is composed from the strongly correlated test statistics.
To obtain a decision on the null hypothesis controlling for a given significance level a permutation test is used. That means the distribution of $T_j$
is determined by using random permutations of variable $j$ that break the relation of the covariate and the response.

Given overall significance level $\alpha$ the significance level for the permutation test that tests splits in one variable is chosen by $\alpha/m$, where $m$ denotes the number of covariates that are available. For the item and variable with the largest value of $T_j$ the permutation test is carried out. If no significant effect is found no splitting is performed. Otherwise for this combination the split point is chosen for which $T_{jc_j}$ had the smallest $p$-value.
Since variable selection is separated from the splitting procedure one could also use alternative criteria for the selection of splits.
If variable, item and split point are selected the model is fitted for this selection yielding estimates $\hat\theta_p, \hat\gamma_{il}, \hat\gamma_{ir}$.

In later steps of the growing of a tree the basic procedure is the same but now one starts from already selected splits.
Let the already built node for item $i$ be characterized by $S_i=\{(c_{ij_1},a_{i1}),\dots,(c_{ij_{k}},a_{ik})\}$,
where $c_{ij_s}$ is the threshold in variable $j_s$ and $a_{is} \in \{0,1\}$ encodes if one is below or above the threshold.
The corresponding node is
\[
\text{node}_i=\prod_{s=1}^{k} I(x_{pj_s} > c_{ij_s})^{a_{is}}(1-I(x_{pj_s} > c_{ij_s}))^{1-a_{is}}.
\]
When considering splits of this node one examines for all variables $j$ and all possible splits the Rasch model with item difficulties
\[
\gamma_{il}\text{node}_i I(x_{pj} \le c_{j}) + \gamma_{ir}\text{node}_i I(x_{pj} > c_{j})
\]
where $c_{j}$ is a split point for variable $j$. The corresponding  null hypothesis is $H_0: \gamma_{il}-\gamma_{ir}=0$, which is tested by test statistic $T_{jc_j}$. Again one first investigates if variable $j$ has an effect by using a permutation test for $T_j= \max_{c_j}T_{jc_j}$ with significance level $\alpha/m$, for the node and variable with the largest value of $T_j$. If a significant effect is found one determines the best split and fits the corresponding model for this split point.
It should be noted that in the fitting step all other parameters of the model, including the person parameters $\theta_p$, are refitted.

After severel splits the tree for item i is defined by terminal nodes $S_{i1},\hdots, S_{iL_i}$ and the predictor of the model can be represented by
\begin{equation}
\eta_{pi} = \theta_p-\text{tr}_i(\xb_p)
= \theta_p-\sum_{\ell=1}^{L_i}{\gamma_{i\ell} \text{ node}_{i\ell}}
\end{equation}
where $\gamma_{i1},\hdots, \gamma_{iL_i}$ denote the item difficulties in the terminal nodes. The algorithm terminates if no permutation test is significant anymore. For those items where no splitting is performed the constant $\text{tr}_i(\xb_p)=\beta_i$, corresponding to the item parameter of the simple Rasch model, is fitted.

%
%
%
%
%
%
%
%
%
%
%

\section{Simulations} \label{sec:simulation}

In this section we  investigate the performance of the fitting procedure in terms of the ability to detect items that show DIF and to estimate the item difficulty parameters in each node. We consider several simulation scenarios where data $y_{pi},\;p=1,\hdots,P,\;i=1,\hdots,I$ were generated according to the binary Rasch model with DIF in some of the items. All the presented results are based on 100 replications.

The following components of the model are the same in each simulation scenario:

\begin{itemize}

\item $P=500$ (number of persons); $\,I=20$  (number of items)
\item $\theta_p \sim N(0,1)$ (person abilities)
\item $\beta_i \sim N(0,1)$  (item difficulties for items without DIF)

\end{itemize}

If item $i$ is assumed to show DIF the corresponding  item difficulties are drawn from a normal distribution. The item difficulties refer to  groups of persons represented by the nodes $S_{i1},\hdots, S_{iL_i}$.

\subsubsection*{Strength of DIF}

In each simulation scenario we generate data with three different strengths of DIF, strong, medium and weak. The strength of DIF in one item $i$ can be measured by the variance of the item parameters $V_i=\text{var}\left(\sum_{\ell}{\gamma_{i\ell} \text{ node}_{i\ell}}\right)$, which for fixed nodes is determined by the parameters $\gamma_{i\ell}$. The average of $V_i$ over the items with DIF is used as a measure of the overall strength of DIF in these items. In all of the simulation scenarios parameters are specified in such a way that for strong DIF the DIF strength is 0.41, for medium DIF the strength is 0.23 and for weak DIF it is 0.10.

\subsubsection*{Mean squared errors}

We compare the estimated coefficients to the true parameters by calculating mean squared errors (MSEs). For the person abilities it is $\frac{1}{P}\sum_{p=1}^{P}({\hat{\theta_p}-\theta_p})^2$ and for the item difficulties it is $\frac{1}{P\cdot I}\sum_{p=1}^{P}\sum_{i=1}^{I}({\hat{tr}_i(\xb_p)-tr_i(\xb_p)})^2$, respectively, averaged over all simulations.

\subsubsection*{Hit rates}

Let each item be characterized by a vector $\deltab_i^T=(\delta_{i1},\hdots,\delta_{im})$, with $\delta_{ij}=1$ if item i has DIF in component $j$ and $\delta_{ij}=0$ otherwise. With indicator function $I(\cdot)$, criteria to judge the identification of items with DIF are:
\begin{itemize}
\item True positive rate on the item level:

$TPR_i=\frac{1}{\#\{i:\deltab_i\neq \mathbf{0}\}}\sum_{i:\deltab_i\neq \mathbf{0}}{I(\hat{\deltab}_i\neq\mathbf{0})}$

\item False positive rate on the item level:

$FPR_i=\frac{1}{\#\{i:\deltab_i=\mathbf{0}\}}\sum_{i:\deltab_i\neq \mathbf{0}}{I(\hat{\deltab}_i\neq\mathbf{0})}$

\item True positive rate for the combination of item and variable:

$TPR_{iv}=\frac{1}{\#\{i,j:\delta_{ij}\neq 0\}}\sum_{i,j:\delta_{ij}\neq 0}{I(\hat{\delta}_{ij}\neq 0)}$

\item False positive rate for the combination of item and variable:

$FPR_{iv}=\frac{1}{\#\{i,j:\delta_{ij}= 0\}}\sum_{i,j:\delta_{ij}\neq 0}{I(\hat{\delta}_{ij}\neq 0)}$
\end{itemize}

\subsection{One single predictor}

In the first simulation scenarios we consider only one predictor $x$ that induces DIF in several items. In this case also traditional methods to detect DIF can be used.

\subsubsection*{Comparison with Alternative Methods}

We will start with a comparison of the proposed method with other established methods for the detection of DIF. Most methods are restricted to the comparison of two or more groups. We consider the Mantel-Haenszel method (MH), the method of logistic regression (Logistic) and Lord's $\chi^2$-test (Lord). An overview of these methods is given in \citet{magisetal:2010} and \citet{magis2011}. For the comparison we use the implementation in the R add-on package \texttt{difR} \citep{difR2013}.

For the comparison of two groups we simulate four items with DIF induced by one binary predictor $x \in \{0,1\}$. For the comparison of multiple groups we simulate DIF with respect to an ordered factor $x \in \{1,\hdots,5\}$. The definition of differences of item difficulties in these groups are given in Table \ref{tab:sim12_structure} for both scenarios. The overall strength of DIF in the four items can be determined by the value of c. Choosing c=1 in the strong setting, c=0.75 in the medium case and c=0.5 in the weak case leads to the DIF strengths as given above.

\begin{table}[!ht]
\centering
\begin{tabularsmall}{lrrrr}
\toprule
&\multicolumn{4}{c}{\bf{Difference of Difficulty}} \\
\bf{Scenario}&Item 1&Item 2&Item 3&Item 4\\
\cline{2-5}
two groups&$1c\cdot I(x=1)$&$-1c\cdot I(x=1)$&$1.5c\cdot I(x=0)$&$-1.5c\cdot I(x=0)$\\
five groups&$1c\cdot I(x>2)$&$-1c\cdot I(x>3)$&$1.5c\cdot I(x>4)$&$-1.5c\cdot I(x>1)$\\
\bottomrule
\end{tabularsmall}
\caption{True simulated differences of item difficulties for the comparison of two or five groups.}
\label{tab:sim12_structure}
\end{table}

\begin{table}[!ht]
\centering
\begin{tabular}{llcccc}
\toprule
&&\multicolumn{2}{c}{\bf{Two groups}}&\multicolumn{2}{c}{\bf{Five groups}}\\
\bf{Method}&&$TPR_i$&$FPR_i$&$TPR_i$&$FPR_i$\\
\cline{2-6}
&strong&0.9975&0.0463&0.9950&0.0581\\\
MH&medium&0.9800&0.0444&0.9125&0.0588\\
&weak&0.8400&0.0450&0.5300&0.0575\\
\cline{2-6}
&strong&0.9975&0.0513&0.9975&0.0656\\
Logistisch&medium&0.9750&0.0506&0.9225&0.0594\\
&weak&0.8375&0.0488&0.5700&0.0600\\
\cline{2-6}
&strong&0.9975&0.0325&0.9850&0.0286\\
Lord&medium&0.9650&0.0325&0.8225&0.0268\\
&weak&0.7900&0.0319&0.3925&0.0300\\
\cline{2-6}
&strong&0.9950&0.0444&0.9900&0.0500\\
IFTrees&medium&0.9625&0.0438&0.9250&0.0581\\
&weak&0.8100&0.0481&0.6100&0.0538\\
\bottomrule
\end{tabular}
\caption{True positiv and false positive rates on the item level for the comparison of two or five groups as average over all 100 replications.}
\label{tab:sim12_results}
\end{table}

The  selection performance for both scenarios is given in Table \ref{tab:sim12_results} for all of the methods. In the case of item focussed trees each permutation test is based on 1000 permutations. Table \ref{tab:sim12_results} shows true positive and false positive rates on the item level as the average over 100 simulations, respectively. It is seen that our proposed method  competes well with the established methods. In the case of two groups true positive and false positive rates are nearly the same for all methods. In the case of five groups and weak DIF the  true positive rates are poor for all of the methods. However, item focussed trees shows the best result  yielding the true positive rate  $0.61$.

\subsubsection*{Continuous Predictor}

The previous simulations showed that  item focussed trees work quite well in pure detection of DIF items when compared to established methods. One of the  advantages of item focussed trees is that the method is not limited to the case of a simple comparison of multiple groups but can also handle a much more complex structure of predictors.

In the following we consider one standard normal distributed predictor $x$ and two items with DIF. We assume a sigmoidal relation between the value of $x$ and the item difficulty of item 1 and 2. The linear predictors are given by
\[
\eta_{p1}=\theta_p-\beta_{1}+c\cdot \text{arctan}(x_p) \quad\text{and}\quad \eta_{p2}=\theta_p-\beta_{2}-c\cdot \text{arctan}(x_p)
\]

\begin{figure}[!ht]
\begin{center}
\includegraphics[width=0.49\textwidth]{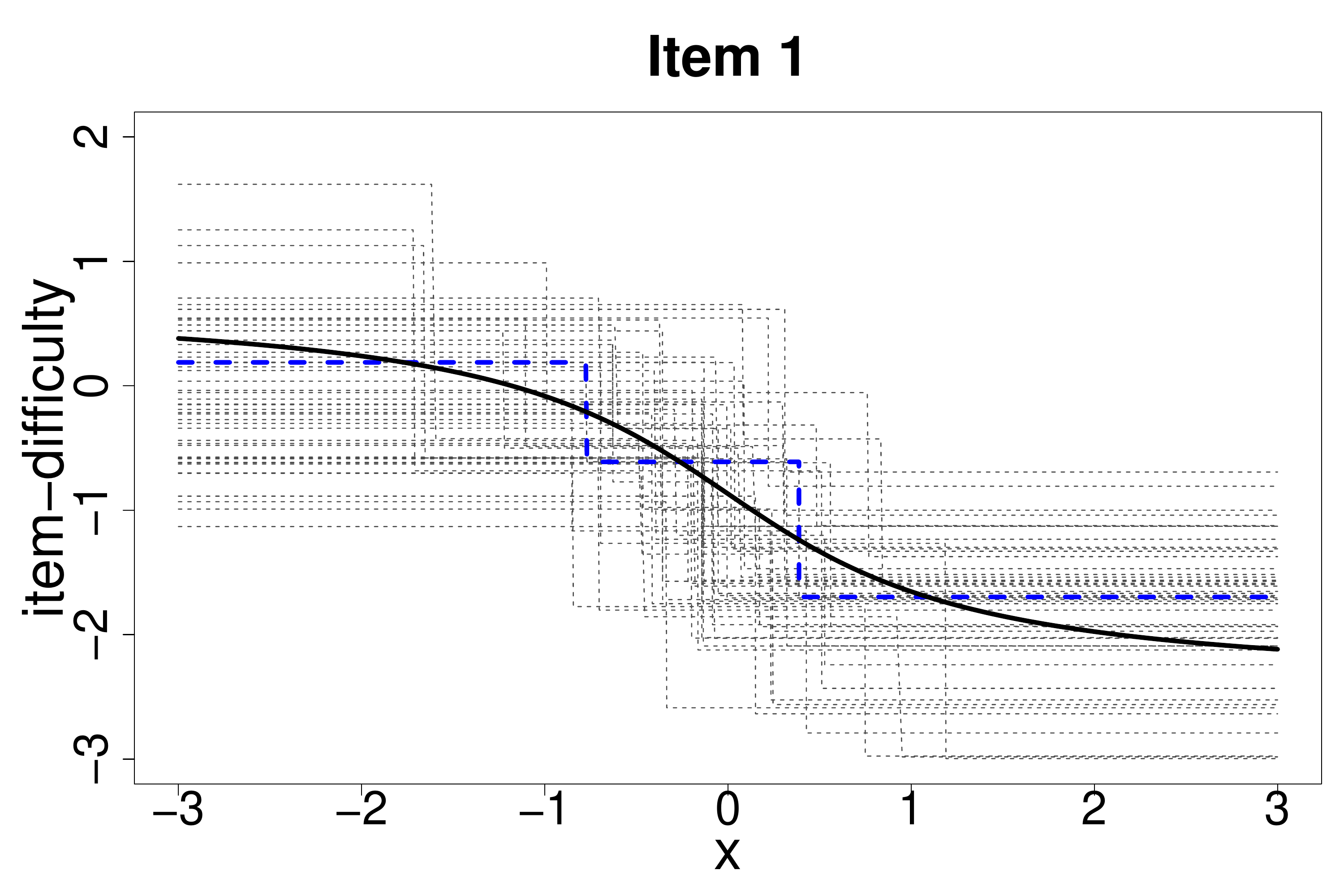}
\includegraphics[width=0.49\textwidth]{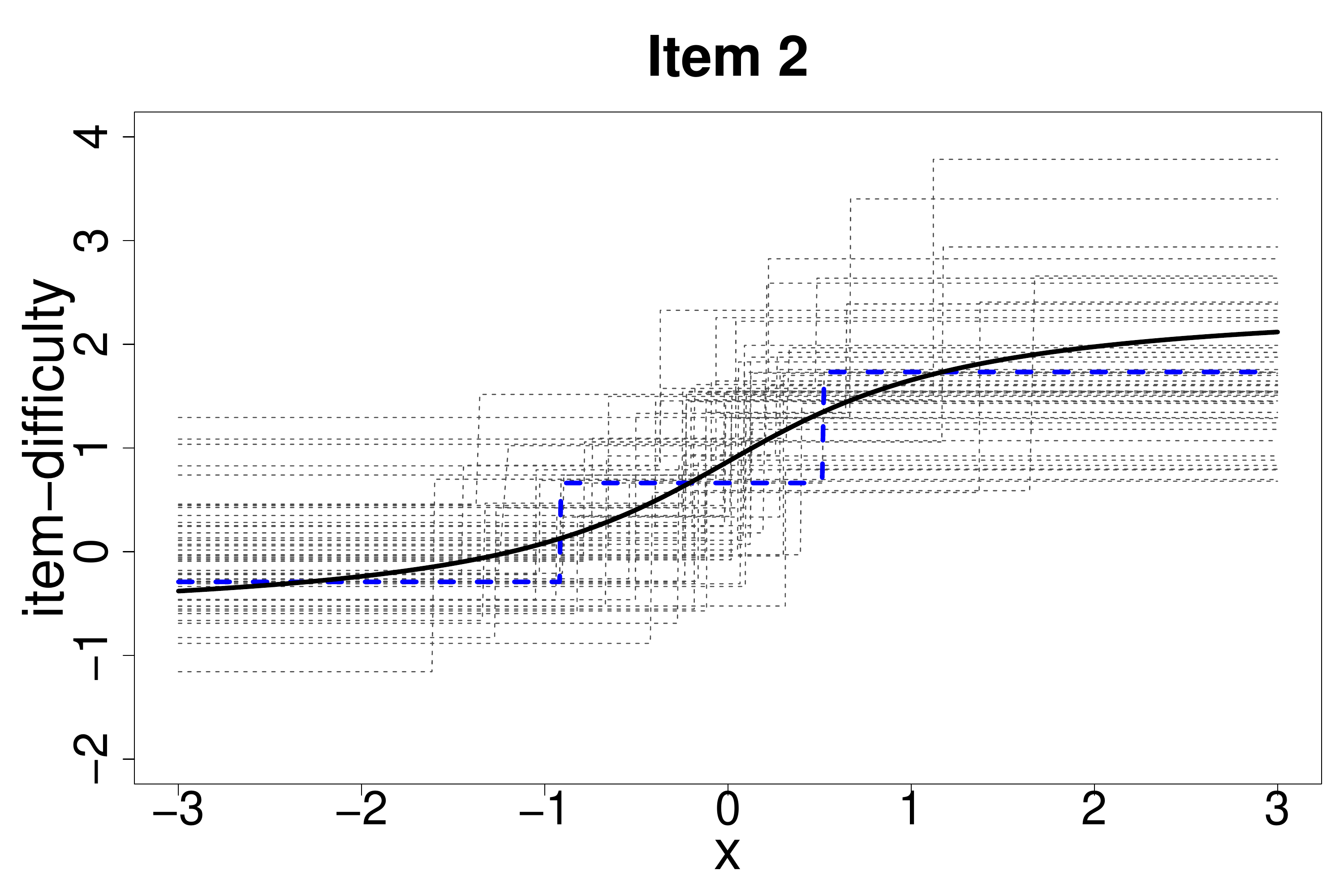}
\end{center}
\caption{True item-difficulties for item 1 and 2 (bold lines) and estimated item difficulties for 50 replications  (dashed lines) of the simulation scenario with one standard normal distributed predictor and strong DIF.}
\label{fig:sim3_stepfunctions}
\end{figure}

For item 1 item difficulties are monotonically decreasing, thus for persons with small $x$ item 1 is harder to solve than for persons with a higher value of $x$. For item 2 item difficulties are monotonically increasing, thus for persons with a small value of $x$ it is easier to solve than for persons with a higher value. The data generating process in this scenario is not determined by step functions but on smooth functions. Therefore the problem is a difficult one for trees, which rely on step functions.   The overall strength of DIF in items 1 and 2 is again determined by a factor $c$. In order to achieve comparable results we use the same values of $c$ as in the previous simulations leading to the same DIF strengths of 0.41 (strong), 0.23 (medium) and 0.10 (weak).

Figure \ref{fig:sim3_stepfunctions} shows the function of the true underlying item difficulties for item 1 and 2 with strong DIF and the estimated step functions for 50 randomly chosen replications of the simulation drawn with dashed lines for $x \in [-3,3]$. It is seen that the estimated step-functions in Figure \ref{fig:sim3_stepfunctions} capture the underlying structure quite well.

\begin{table}[!ht]
\begin{center}
\begin{tabular}{llcccc}
\toprule
\bf{Scenario}&&MSE persons&MSE items&$TPR_i$&$FPR_i$\\
\cline{2-6}
&strong&0.4511&0.1585&1.0000&0.0411\\
3&medium&0.4378&0.1533&0.9750&0.0439\\
&weak&0.4257&0.1439&0.7550&0.0439\\
\bottomrule
\end{tabular}
\end{center}
\caption{Estimated MSEs, true positive rates and false positive rates for the simulation scenario with one standard normal distributed predictor as average over 100 simulations.}
\label{tab:sim3_results}
\end{table}

Estimated MSEs of person-parameters $\theta_p$ and item-parameters $tr_i(\xb_p)$ as well as true positive and false positive rates on the item level  averaged over all simulations are given in Table \ref{tab:sim3_results}. Again all permutation tests are based on  1000 permutations. In the case of one single predictor $x$ vector $\deltab_i$ only has one element, so true positive and false positive rates for the combination of item and variable correspond to those on the item level.

Similar to the results in Table \ref{tab:sim12_results} true positive rates in Table \ref{tab:sim3_results} are very high even in the case of weak DIF. False positive rates are smaller than $0.05$ so the global significance
level holds. As was to be expected MSEs of person parameters and item parameters slightly grow with increasing strength of DIF.

\begin{figure}[!ht]
\begin{center}
\includegraphics[width=0.49\textwidth]{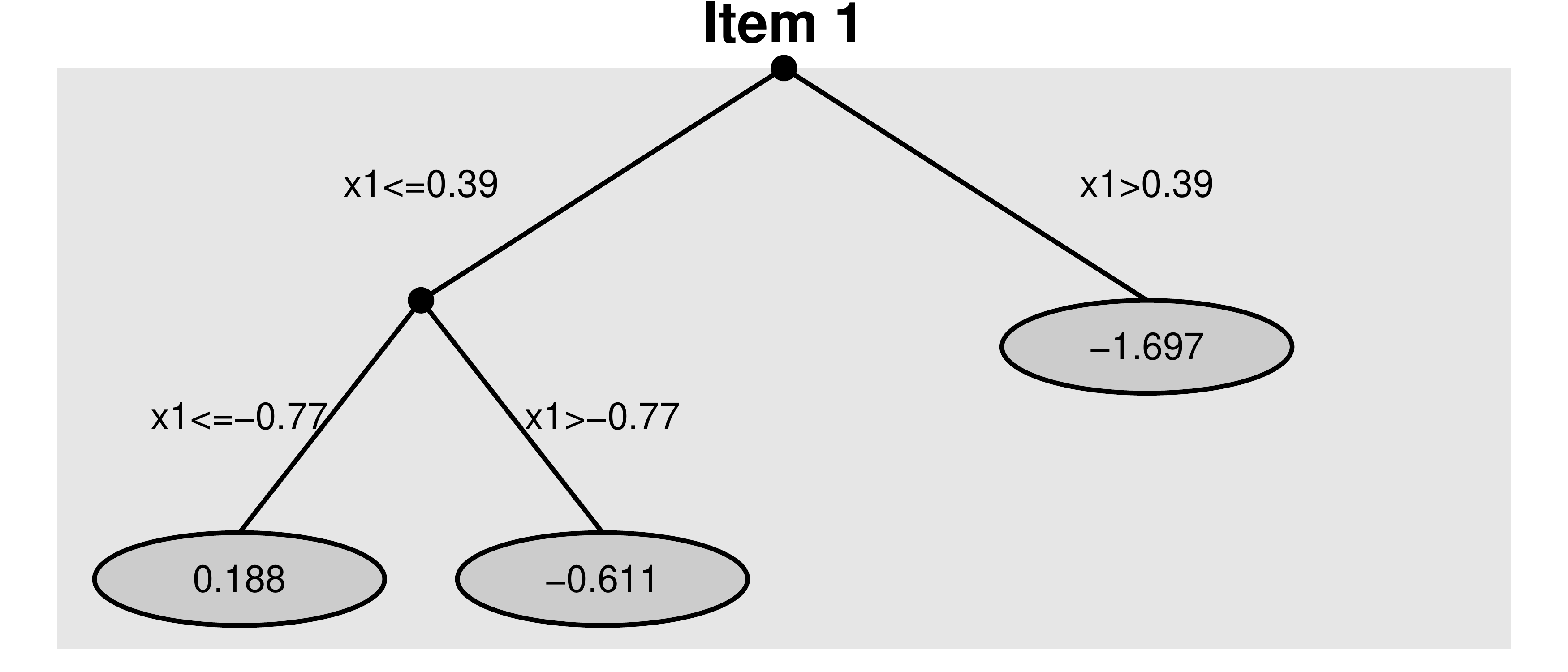}
\includegraphics[width=0.49\textwidth]{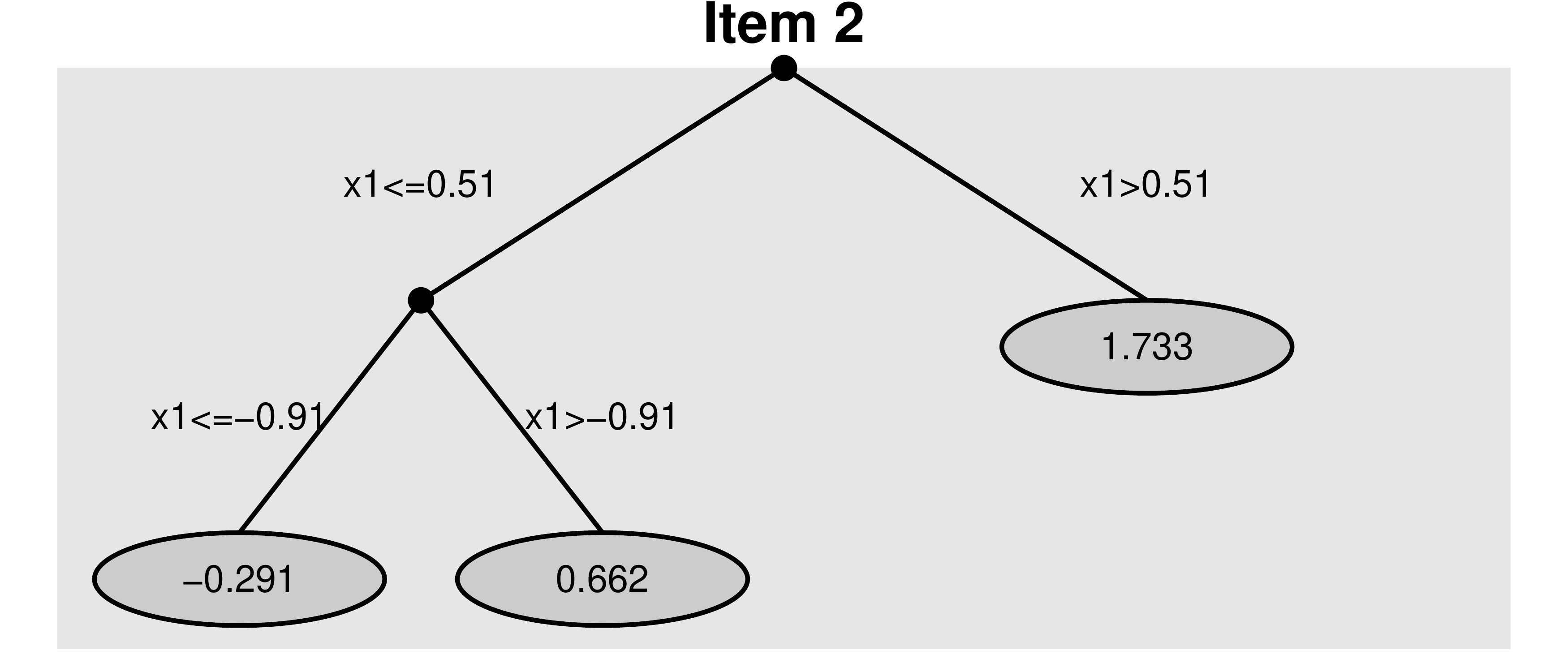}
\end{center}
\caption{Trees for item 1 and 2 for one estimation of the simulation with one standard normal distributed predictor and strong DIF. Estimated item difficulties are given in each leaf of the trees.}
\label{fig:sim3_trees}
\end{figure}

Single estimation results can also be visualized as tree. Figure \ref{fig:sim3_trees} shows the resulting trees for item 1 and 2 for one exemplary replication of the simulation with strong DIF. The estimated item difficulties are given in each leaf of the trees. In this example two splits are performed for both items. Because of small differences of item difficulties at the borders the algorithm does not perform more splits. A tree with 2 splits or 3 leafs corresponds to a estimated function with 2 steps. The corresponding step functions are marked by  dashed lines in Figure \ref{fig:sim3_stepfunctions}.

The simulation scenario shows that the proposed method is not only able to find relevant DIF items but also to detect complex, especially not linear, structures of DIF. Also in terms of estimation accuracy the algorithm  performs quite well.

\subsection{Several predictors}

In the following simulations we consider data with four predictors $x1,\hdots,x4$ that potentially induce DIF in 4 out of 20 items. The distributions of the four predictors are
\[
x1,\,x3 \sim B(1,0.5) \quad \text{and} \quad x2,\,x4 \sim N(0,1).
\]

\begin{table}[!ht]
\begin{center}
\begin{tabularsmall}{lrrr}
\toprule
&\multicolumn{3}{c}{\bf{Difference of Difficulty}} \\
\bf{Item}&Scenario 4&Scenario 5&Scenario 6\\
\cline{2-4}
1&$1c\cdot I(x1=1)$&$1c\cdot I(x2>0.1)$&$0.75c\cdot I(x1=1)+0.75c\cdot I(x2>0.1)$\\
2&$-1c\cdot I(x1=1)$&$-1c\cdot I(x2>0.1)$&$-0.75c\cdot I(x1=1)-0.75c\cdot I(x2>0.1)$\\
3&$1.5c\cdot I(x3=1)$&$1.5c\cdot I(x4>-0.1)$&$0.8c\cdot I(x3=1)+0.8c\cdot I(x4>-0.1)$\\
4&$-1.5c\cdot I(x3=1)$&$-1.5c\cdot I(x4>-0.1)$&$-0.8c\cdot I(x3=1)-0.8c\cdot I(x4>-0.1)$\\
\bottomrule
\end{tabularsmall}
\end{center}
\caption{True simulated differences of item difficulties for the three simulation scenarios with four predictors.}
\label{tab:sim46_structure}
\end{table}

We consider three simulation scenarios with different structures of DIF with respect to Items 1, 2, 3 and 4. Differences of item difficulties are defined as given in Table \ref{tab:sim46_structure}. In scenario 2 DIF occurs in the binary components $x1$ and $x3$, in scenario 3 DIF occurs in the continous components $x2$ and $x4$ and in scenario 4 it is a more complex structure with DIF in a combination of binary and normal distributed variables. The overall strength of DIF in the four items again depends on the value of c and is set the same as in the previous scenarios.

\begin{figure}[!ht]
\begin{center}
\includegraphics[width=0.4\textwidth]{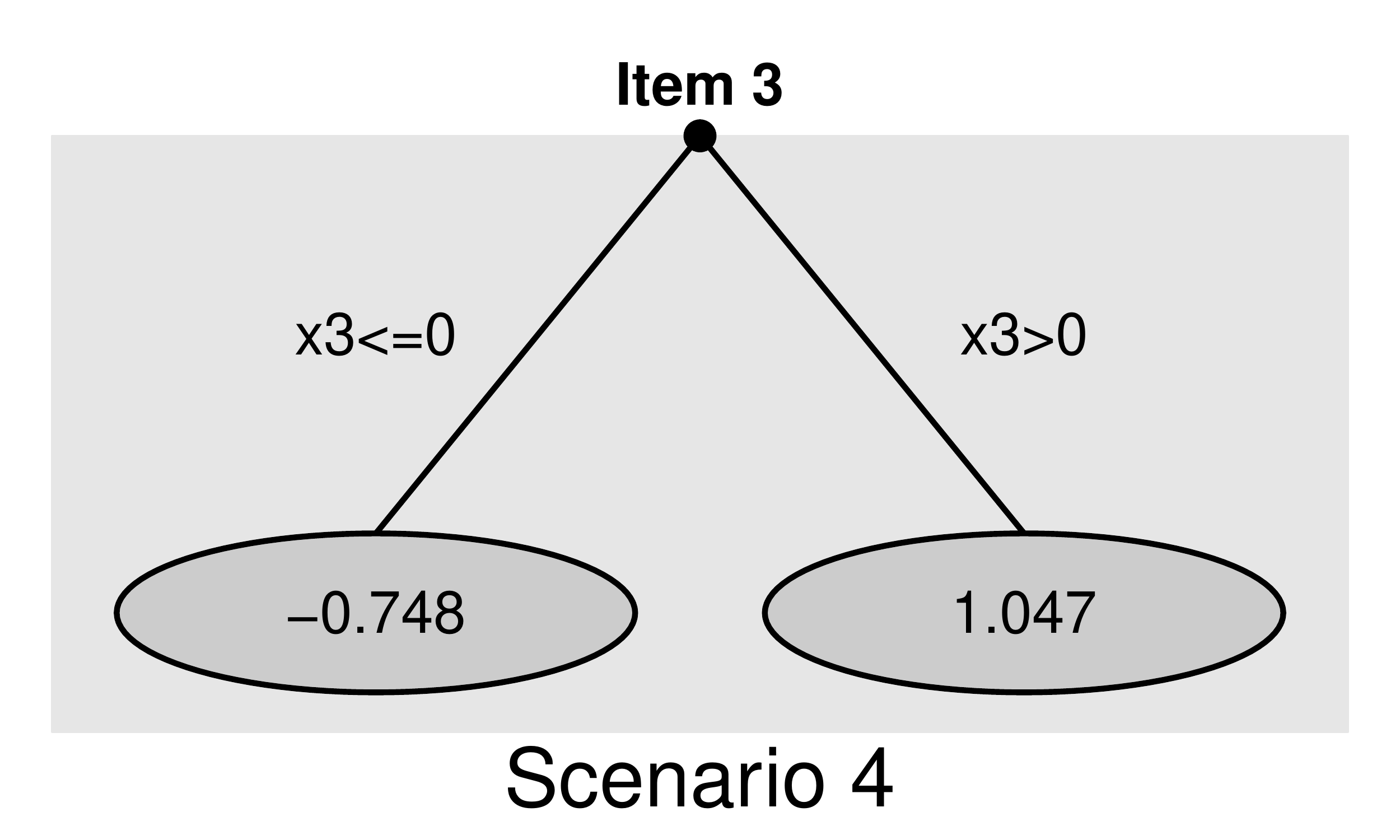}
\includegraphics[width=0.4\textwidth]{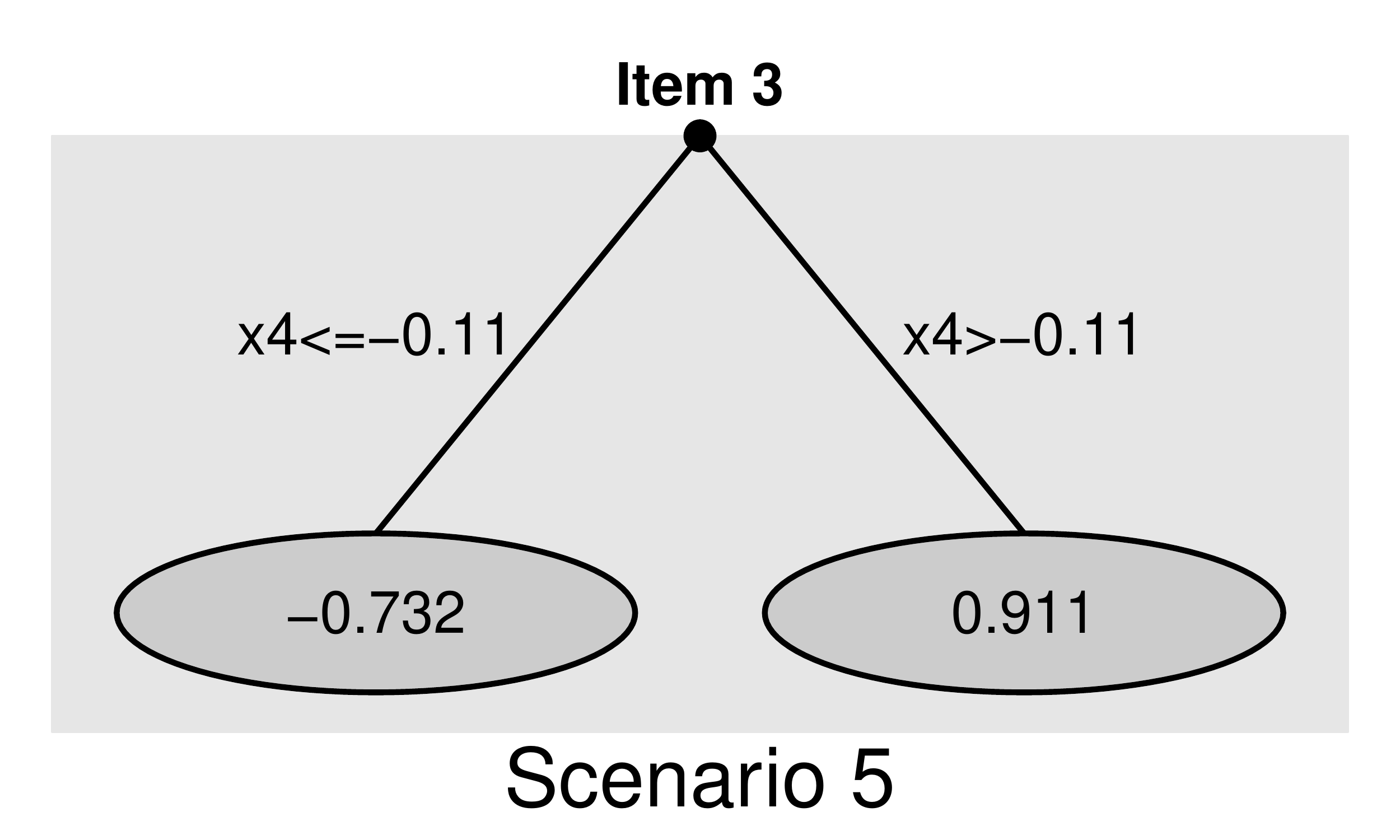}
\includegraphics[width=0.8\textwidth]{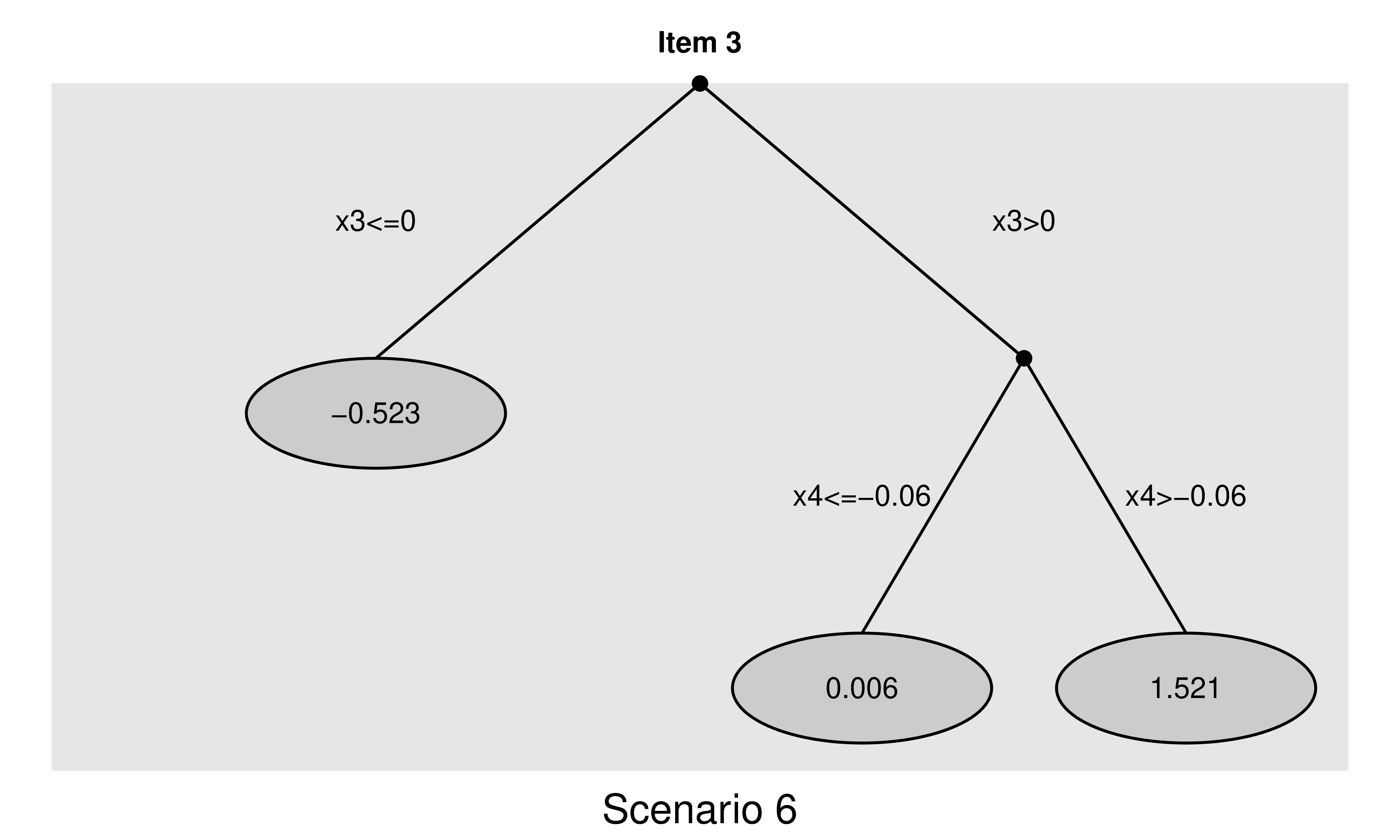}
\end{center}
\caption{Exemplary estimation results for the strong setting of simulation scenario 4, 5 and 6 with four predictors.  Estimated item difficulties are given in each leaf of the trees.}
\label{fig:sim46_trees}
\end{figure}

Figure \ref{fig:sim46_trees} shows one exemplary estimation result of item 3 for each scenario with strong DIF where the true underlying tree structure is detected. The estimated item difficuties are given in each leaf of the trees. The true item parameters for item 3 of the two groups in scenario 4 and 5 are $-0.68$ and $0.82$. In scenario 6 they are $-0.68$, $0.12$ and $0.92$. As for all other simulations estimated values are  close to the true ones. True and estimated split points of scenario 5 and 6 regarding to the standard normal variable $x4$ do not differ very much for the exemplary trees in Figure \ref{fig:sim46_trees}. Due to the data generating process they are clearly not exactly the same. For the binary variable $x3$ there is only one possible split.

An overview of the simulation results based on 100 replications is given in Table \ref{tab:sim46_results}. MSEs of person-parameters $\theta_p$ and item-parameters $tr_i(\xb_p)$, true-positive and false-positive rates on the item level as well as for the combination of items and variables are summarized for the three scenarios and each strength of DIF. All permutation tests are again based on 1000 permutations. To account for the four covariates in the model the local significance level for one test is $0.05/4$.

\begin{table}[!ht]
\begin{center}
\begin{tabular}{llcccccc}
\toprule
&&\multicolumn{2}{c}{\bf{MSE}}&\multicolumn{2}{c}{\bf{true positive}}&\multicolumn{2}{c}{\bf{false positive}}\\
\bf{Scenario}&&persons&items&$TPR_i$&$TPR_{iv}$&$FPR_i$&$FPR_{iv}$\\
\cline{2-8}
&strong&0.4253&0.1336&0.9825&0.9825&0.0269&0.0089\\
4&medium&0.4034&0.1299&0.8375&0.8350&0.0270&0.0084\\
&weak&0.4056&0.1272&0.4975&0.4900&0.0263&0.0074\\
\cline{2-8}
&strong&0.4176&0.1583&0.9625&0.9625&0.0275&0.0087\\
5&medium&0.4111&0.1474&0.8375&0.8350&0.0313&0.0084\\
&weak&0.4174&0.1649&0.5300&0.5275&0.0263&0.0064\\
\cline{2-8}
&strong&0.4207&0.1459&0.9950&0.6800&0.0244&0.0072\\
6&medium&0.4105&0.1392&0.8700&0.5250&0.0269&0.0076\\
&weak&0.4086&0.1423&0.4325&0.2300&0.0310&0.0076\\
\bottomrule
\end{tabular}
\end{center}
\caption{Simulation results for simulation scenarios 4, 5 und 6 with four predictors as the average over 100 simulations.}
\label{tab:sim46_results}
\end{table}

It is seen that MSEs of person parameters tend to  grow with increasing strength of DIF but are quite stable over all simulations. Hence estimation accuracy is  affected not too much by variable and DIF structure. MSEs of item parameters are about the same as in Table \ref{tab:sim3_results} but do not differ systematically.
True positive rates on the item level are very high for medium and strong DIF for each of the three scenarios. Detection of relevant DIF inducing items works well in this settings. In the weak settings only about half of the items with DIF are identified.
In scenario 6 DIF is affected by two variables. Here true positive rates for the combination of item and variables are clearly smaller than for scenario 4 and 5. Even for strong DIF the hit rate for item and variable is only about $0.68$.
False positive rates are very small across all simulations, in particular the global significance level holds. At most one item without DIF is misleadingly identified as DIF item or one split with regard to a variable that was not inducing DIF is executed during estimation.

\begin{figure}[t]
\centering
\includegraphics[width=1\textwidth]{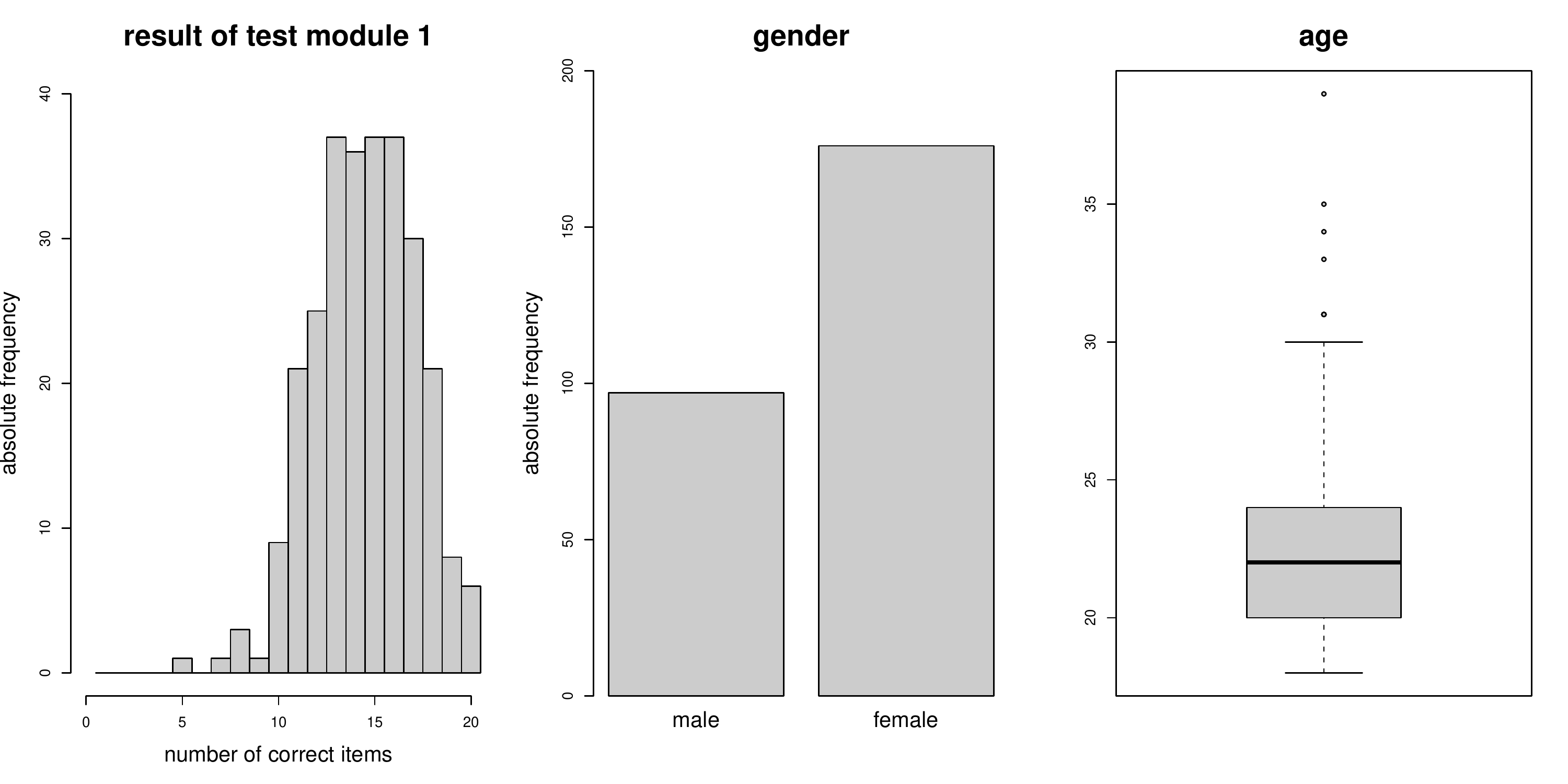}
\caption{Graphical representation of the results of the first module (items 1 to 20) of the I-S-T 2000 R (left) and the distribution of the two covariates in the analyzed data.}
\label{fig:Deskriptiv_IST}
\end{figure}

\begin{figure}[t]
\begin{center}
\includegraphics[width=0.4\textwidth]{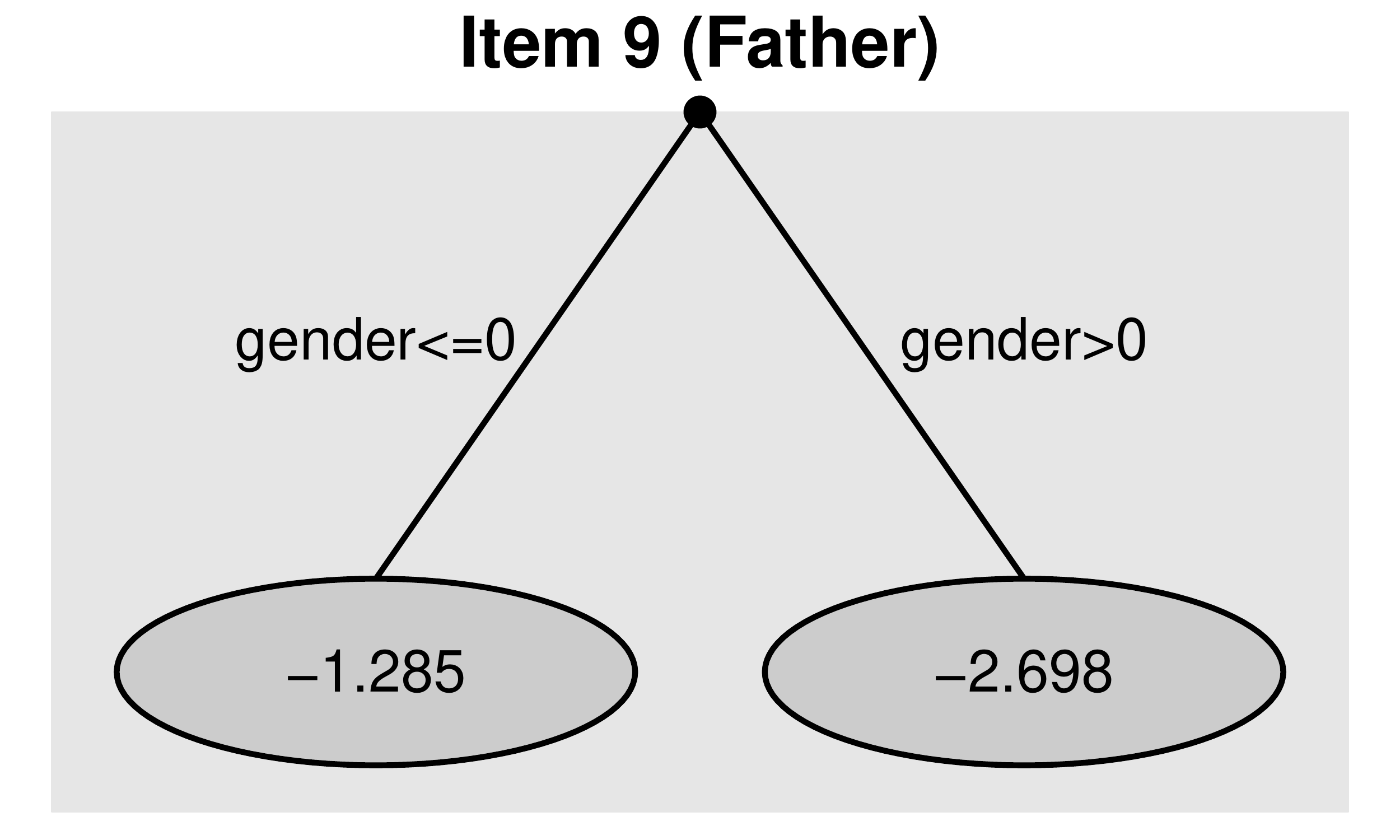}
\includegraphics[width=0.4\textwidth]{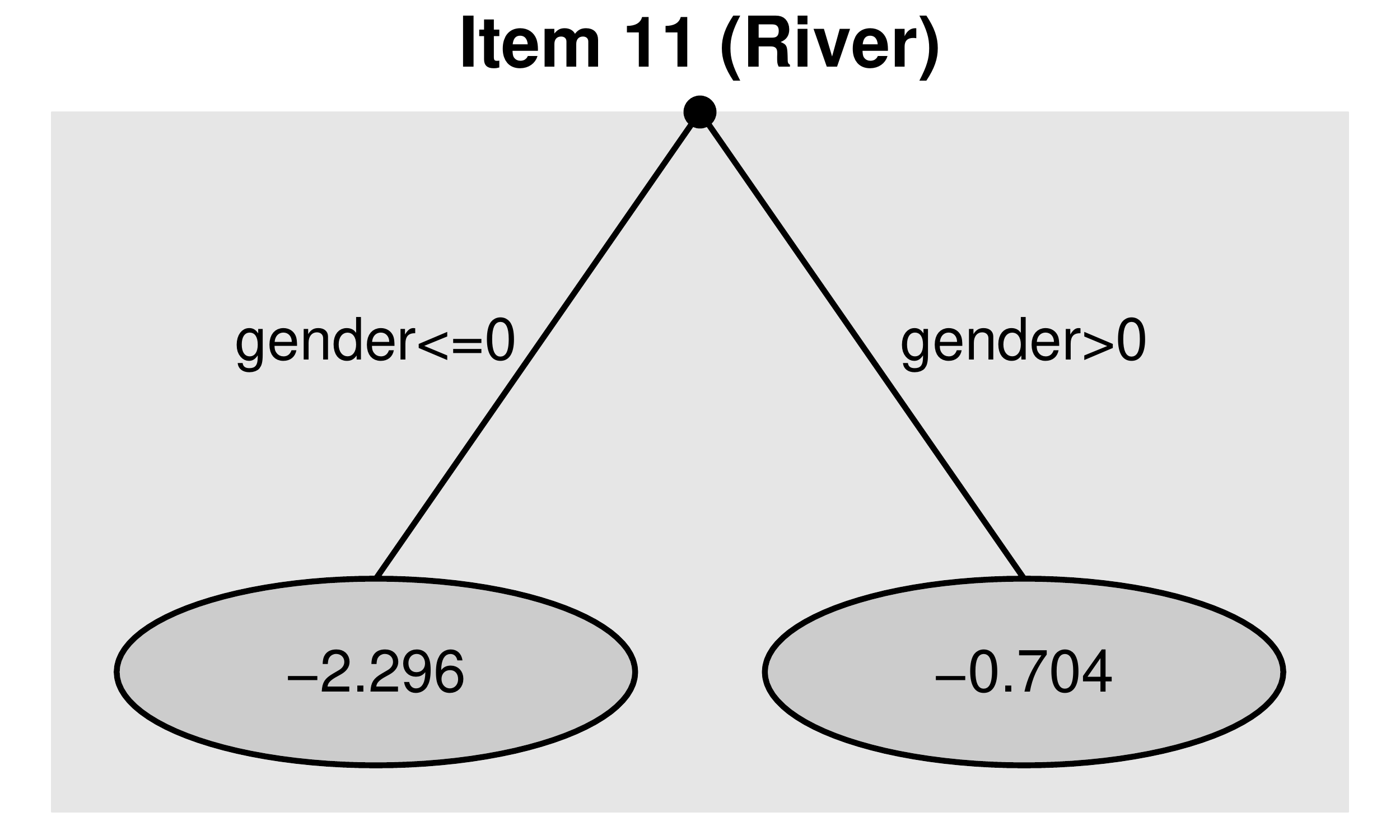}
\includegraphics[width=0.8\textwidth]{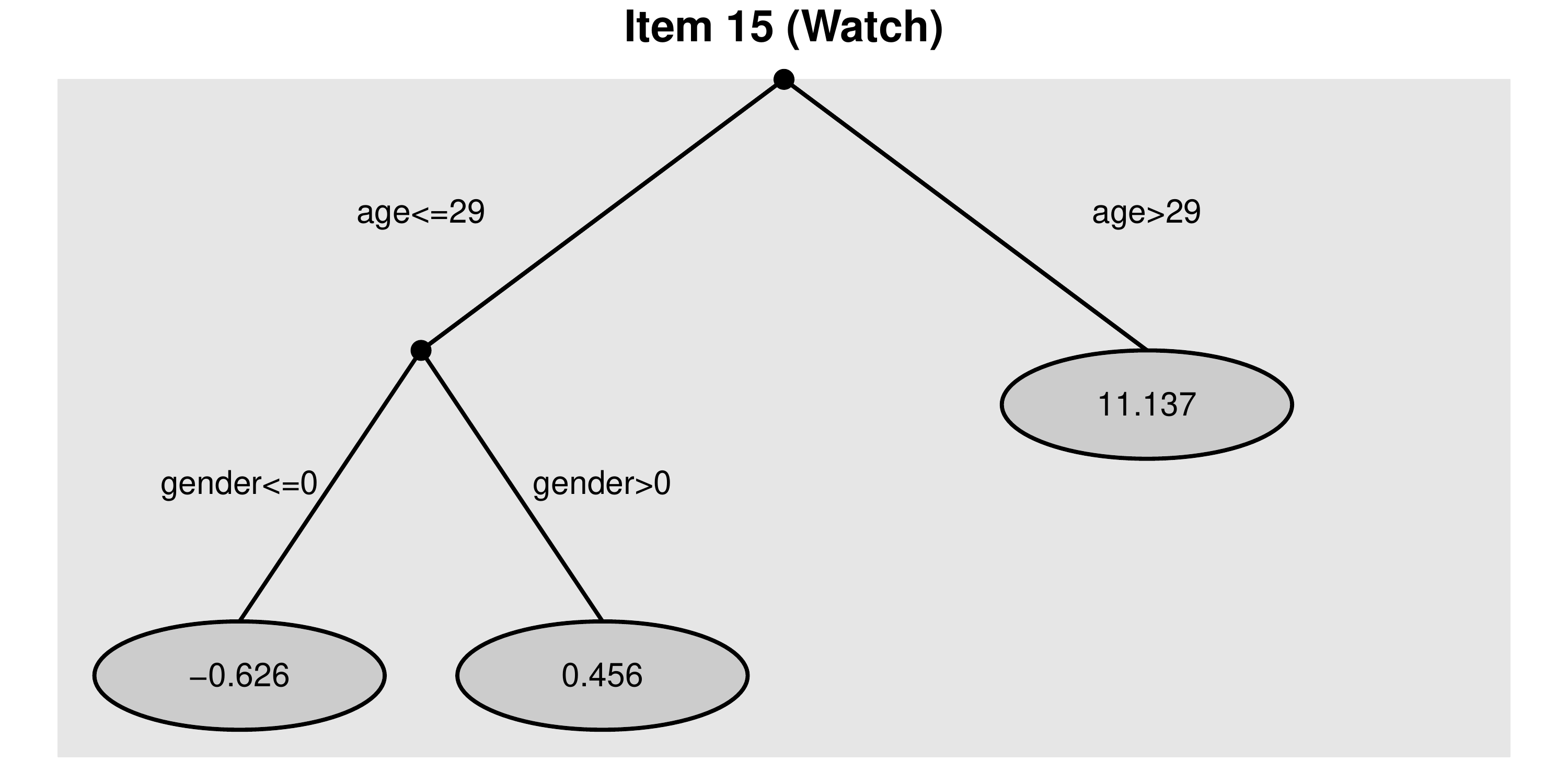}
\end{center}
\caption{Trees for Items 9, 11 and 15 of the I-S-T 2000 R. Estimated item difficulties are given in
each leaf of the trees.}
\label{fig:Tree_91115}
\end{figure}

\section{Further Application} \label{sec:application}
As second application we consider data of the Intelligence-Structure-Test 2000 R (I-S-T 2000 R) see, for example, \citet{Amthauer1999, Amthauer2001}. The present study was carried out by the Department of Education  of the Ludwig-Maximilians University in Munich \citep{Buhner2006}. The test was conducted at the Phillips University in Marburg. For our analysis we use data from 273 students from 40 different subject areas. The I-S-T 2000 R consists of 9 modules with 20 items each. The first module (items 1 to 20) is about the completion of sentence and asks for sentences where one word is missing. There are five possible solutions for each sentence. The respondent is asked to choose the word that completes the sentences correctly.

To test for DIF in these items we incorporate the covariates gender (male: 0, female: 1) and age. The distribution of the two covariates and the test result for items 1 to 20 are displayed in Figure \ref{fig:Deskriptiv_IST}. There are 97 male and 176 female students with age ranging from 18 to 39. The student with the worst result had only 5 correct answers, whereas six students answer all 20 tasks of module 1 correctly.

Using item focussed recursive partitioning results in 3 of 20 items showing DIF. The algorithm executed only four splits before stopping ($\alpha=0.05$). All permutation tests were based on 1000 permutations. Both covariates gender and age are at least once used for splitting and therefore both covariates are included in the model. The three items that were identified as DIF items are the following (correct answers are marked in bold):

\begin{itemize}
\item[9:] Fathers are ...? (more) experienced than their sons.

a) always$\quad$ {\bf b) usually}$\quad$ c) much$\quad$  d) less$\quad$  e) fundamentally
\item[11:] Every river has ...?

a) fishes$\quad$ b) bridges$\quad$ c) ships$\quad$  {\bf d) gradients}$\quad$  e) rapids
\item[15:] A watch always needs (a) ...?

a) battery$\quad$ b) case$\quad$ c) numbers$\quad$  {\bf d) energy}$\quad$  e) hands
\end{itemize}

The resulting trees for items 9, 11 and 15 are shown in Figure \ref{fig:Tree_91115}. Items 9 and 11 show DIF only for gender. The estimated item difficulties show that item 9, which relates to social relations, is easier for females (gender=1) and item 11, which relates to natural sciences, is easier for males (gender=0). Item 15, which relates to technics, is very difficult for all students who are comparably old (age $>$ 29) while for younger students (age $\leq$ 29) it is distinguished between males with an item difficulty of $-0.626$ and females with a larger item difficulty of $0.456$.

The item difficulty of item 15 for students older than 29 given in Figure \ref{fig:Tree_91115} is 11.137. This corresponds to probabilty 1 for solving the item. In fact no student in the sample, who was older than 29, answered item 15 correctly. Thus, when searching for the optimal split, the split regarding age and threshold 29 is obviously the best choice. Splitting in this case leads to a pure node with all responses having value 0. A maximum likelihood estimate for the item difficulty in this node does not exist as it tends to infinity. In order to guarantee the existence of all estimates we added a small ridge penalty on the item parameters that ensures that an estimate exists.

\section{Concluding Remarks} \label{sec:remarks}

Item focussed recursive partitioning is a modelling tool that allows for simultaneaous detection of items and variables that are responsible for DIF. In particular when several covariates on different scales are available as potentially DIF inducing variables it is an efficient and flexible tool for DIF investigations.

Simulation results show that the proposed fitting procedure works quite well in terms of selection performance as well as in terms of estimation accuracy. The results shown here are obtained by an R program that is available from the authors and will soon be available on CRAN.

\bibliography{literatur}

\end{document}